\documentclass[journal,10pt]{IEEEtran}
\usepackage{cite}
\usepackage{amsmath,amssymb,amsfonts}
\usepackage{algorithm, algorithmic}
\usepackage{graphicx}
\usepackage{subfigure}
\usepackage{textcomp}
\usepackage{multirow}
\usepackage{stfloats}
\usepackage{url}
\usepackage{color}
\usepackage{bm}
\usepackage{makecell}
\usepackage{flushend}
\usepackage{float}
\usepackage{threeparttable}
\usepackage{setspace}
\usepackage{upgreek}
\usepackage{bbm}
\usepackage{pifont}

\usepackage[colorlinks,
linkcolor=blue,
anchorcolor=blue,
citecolor=blue, 
urlcolor=black,
]{hyperref}

\newcommand{\nop}[1]{}
\allowdisplaybreaks[4]

\begin{document}
	
	\title {Hybrid RIS-Aided Digital Over-the-Air Computing for Edge AI Inference: Joint Feature Quantization and Active-Passive Beamforming Design
	}
	
	\author{~Yang~Fu,~Peng~Qin,~\IEEEmembership{Member,~IEEE}, Liming Chen, Xianchao Zhang, and Yifei Wang
		
		\thanks{This work was supported in part by 62201212, 62271201, F2025502015, 2025MS008, SEPRI-K24B018, 2025ZD0804700 and MPIS202409. 
			%This work was supported in part by the National Natural Science Foundation of China under Grant 62201212, 62271201, in part by the Natural Science Foundation of Hebei Province under Grant F2025502015, in part by the Fundamental Research Funds for the Central Universities under Grant 2025MS008, and in part by the Science and Technology Projects of China Southern Power Grid under Grant SEPRI-K24B018.
			\textit{(Corresponding author: Peng Qin)}}% <-this % stops a space
		\thanks{
			Yang Fu, Peng Qin, Yifei Wang are with the State Key Laboratory of Alternate Electrical Power System with Renewable Energy Sources, School of Electrical and Electronic Engineering, North China Electric Power University, Beijing, 102206, China (e-mail: qinpeng@ncepu.edu.cn).}
		\thanks{
			Liming Chen is with the Electric Power Research Institute, China Southern Power Grid, Guangzhou, 510663, China.}
		\thanks{
				Xianchao Zhang is with the Provincial Key Laboratory of Multimodal Perceiving and Intelligent Systems, Jiaxing University, Jiaxing, 314001, China.}
		
		}

	\markboth{IEEE Transactions on Communications}
	{}
	%{Shell \MakeLowercase{\textit{et al.}}: Journals}

	\maketitle

	\begin{abstract}	
    The vision of 6G networks aims to enable edge inference by leveraging ubiquitously deployed artificial intelligence (AI) models, facilitating intelligent environmental perception for a wide range of applications. A critical operation in edge inference is for an edge node (EN) to aggregate multi-view sensory features extracted by distributed agents, thereby boosting perception accuracy. Over-the-air computing (AirComp) emerges as a promising technique for rapid feature aggregation by exploiting the waveform superposition property of analog-modulated signals, which is, however, incompatible with existing digital communication systems. Meanwhile, hybrid reconfigurable intelligent surface (RIS), a novel RIS architecture capable of simultaneous signal amplification and reflection, exhibits potential for enhancing AirComp. Therefore, this paper proposes a Hybrid RIS-aided Digital AirComp (HRD-AirComp) scheme, which employs vector quantization to map high-dimensional features into discrete codewords that are digitally modulated into symbols for wireless transmission. By judiciously adjusting the AirComp transceivers and hybrid RIS reflection to control signal superposition across agents, the EN can estimate the aggregated features from the received signals. To endow HRD-AirComp with a task-oriented design principle, we derive a surrogate function for inference accuracy that characterizes the impact of feature quantization and over-the-air aggregation. Based on this surrogate, we formulate an optimization problem targeting inference accuracy maximization, and develop an efficient algorithm to jointly optimize the quantization bit allocation, agent transmission coefficients, EN receiving beamforming, and hybrid RIS reflection beamforming. Experimental results demonstrate that the proposed HRD-AirComp outperforms state-of-the-art digital AirComp baselines in terms of both inference accuracy and uncertainty, achieving performance close to the idealized case with perfect feature aggregation. 
	\end{abstract}


	\IEEEpeerreviewmaketitle
	
	\section{Introduction}
	
	\IEEEPARstart{A}{}wide array of killer applications envisioned for 6G, including humanoid robots, autonomous driving, and low-altitude economy, necessitate intelligent environmental perception to facilitate real-time monitoring and high-precision control of dynamic physical entities \cite{1,2}. To fulfill these stringent requirements, IMT-2030 has identified two pivotal network capabilities: sensing and artificial intelligence (AI) \cite{3}. The former involves the fusion of multi-view sensory data acquired through collaborative devices, while the latter aims to enable the ubiquitous deployment of advanced AI models. The natural integration of these two capabilities at the network edge gives rise to a novel paradigm termed edge inference, referring to the intelligent processing of sensory data via well-trained AI models deployed across edge infrastructure and distributed devices \cite{4}. 
	
	In typical edge inference systems, distributed devices outfitted with sensors (e.g., cameras and radars) and hardware accelerators (e.g., GPUs), termed as agents, collect sensory data from their individual viewpoints and perform local feature extraction using lightweight AI models. The extracted features are transmitted to an edge node (EN), where they are aggregated and then processed by a computation-intensive edge model to complete the inference task \cite{5}. However, the aforementioned feature aggregation, serving as a critical operation in edge inference by harnessing multi-view information to enhance perception accuracy, suffers from a communication bottleneck. This stems from the concurrent transmission of high-dimensional feature vectors over resource-limited wireless channels, resulting in substantial aggregation latency that fails to meet the timeliness demands of inference tasks. To address this bottleneck, over-the-air computing (AirComp) has emerged as a promising solution by exploiting the waveform superposition property to enable rapid aggregation of simultaneously transmitted data streams \cite{6,7,Du2024Distributed}. This appealing characteristic has spurred growing research interest in the seamless integration of AirComp into edge inference systems.
	
	\subsection{Related Works}
	
    \textit{1) AirComp-Enabled Edge Inference:} In this context, AirComp facilitates the over-the-air aggregation of local features transmitted by agents, enabling the EN to reconstruct global features directly from the superimposed signals. Unlike traditional AirComp that primarily focuses on minimizing computation error, the design objective in AirComp-enabled edge inference shifts toward boosting the end-to-end (E2E) inference performance in a task-oriented manner \cite{8}. This inspires a series of studies, ranging from inference performance analysis to task-oriented transceiver design. Specifically, based on the assumption that the global features obey a Gaussian mixture distribution, literature \cite{9} established a theoretical framework for analyzing the scaling law between inference uncertainty and receiving signal-to-noise ratio of AirComp. Work in \cite{10} introduced the concept of AirComp mutual information (MI), which was shown to be generalized to the inference accuracy metric in edge inference systems. To maximize the AirComp MI, the authors developed a majorization-minimization-based beamforming method. \cite{11} investigated a broadband AirComp scenario where different dimensions of local feature were simultaneously aggregated at multiple subcarriers, thereby improving communication efficiency. Subsequently, subcarrier allocation and power control were jointly optimized to reduce the distortion of reconstructed global features. The authors of \cite{12} further considered a multi-task edge inference system, in which MIMO beamforming was optimized to suppress inter-task interference and facilitate accurate AirComp-based feature aggregation. Meanwhile, batching was designed to strategically assemble the global features of different tasks for parallel inference. It is worth noting that the aforementioned works all resort to analog AirComp, wherein each agent linearly modulates local feature values onto the amplitude of transmission signals, thereby making signal superimposition equivalent to feature aggregation. Nevertheless, the prevalent adoption of digital modulation in practical wireless communication systems poses significant challenges to the deployment of analog AirComp on existing hardware. Additionally, the aggregation of analog-modulated features is inherently susceptible to channel impairments, which degrades the reliability of edge inference.

    \textit{2) Digital AirComp Designs:} To overcome the limitations of analog AirComp, several early studies have explored digital AirComp designs. In this line of research, quantization serves as a fundamental operation that transforms continuous-valued data into a finite set of discrete levels, which are then mapped to digitally modulated symbols for over-the-air aggregation. Reference \cite{13} proposed a one-bit digital aggregation scheme, in which each agent quantized its data to a single bit using a sign function, enabling direct mapping to BPSK or QAM symbols. Afterwards, a majority-vote-based decoder was employed at the EN to recover the aggregation result via AirComp. The single bit aggregation was further elaborated in \cite{14}, with additionally design of digital channel coding, thereby addressing the phase asynchrony issue inherent in AirComp. To eliminate the need of precise channel estimations in AirComp, \cite{15} allowed the agents to transmit the quantized versions of analog data on two subcarrier frequencies, then the EN detected the aggregation result via non-coherence receiver. Literature \cite{16} contributed to improving the reliability of AirComp through bit-slicing, where the bit sequence quantizing an analog value was partitioned into segments, each of which was modulated using a constellation with small size. Work in \cite{17} developed a massive digital AirComp scheme, where a large number of agents could freely transmit their signals without the permission of the EN. Under this setup, the authors designed a quantitation codebook for mapping the agents’ analog values into codewords, and proposed a signal detection algorithm to estimate the number of agents applying each codeword. However, most existing digital AirComp designs are limited to scalar quantization, wherein each data value is independently mapped to a codeword. This incurs substantial communication overhead when applied to high-dimensional feature vectors, which often consist of tens of thousands of elements. Incorporating efficient vector quantization techniques into digital AirComp is still an open question that warrants further investigation. Moreover, in edge inference systems, feature elements typically exhibit unequal importance \cite{18,19}, yet this characteristic is neglected in existing quantization schemes. Consequently, task-oriented digital AirComp designs that explicitly aim to enhance inference performance remain largely unexplored. 
    
    \textit{3) Reconfigurable Intelligent Surface (RIS)-Aided AirComp:} Conventional AirComp primarily relies on transceiver design to mitigate the adverse effects of wireless transmission, leaving the propagation environment, characterized by channel fading and blockages, largely uncontrollable. Motivated by recent advances in RIS, a number of studies have explored the additional degrees of freedom (DoFs) offered by RIS to proactively reconfigure wireless channels, thereby enhancing AirComp performance. Paper \cite{20} considered a passive RIS-aided AirComp scenario, and developed a robust design approach for jointly optimizing AirComp transceiver and RIS reflection beamforming, thereby minimizing the computation error under environmental imperfections. In \cite{21}, a pair of RISs were deployed at the agent and EN sides, respectively, for further reducing the computation error of AirComp. Thereafter, the passive beamforming matrices of the two RISs were carefully optimized to coordinate the single- and double-reflection links. More recently, the RIS-aided AirComp framework has been extended to incorporate various advanced RIS architectures, such as active RIS \cite{22}, simultaneously transmitting and reflecting (STAR) RIS \cite{23}, and multi-functional RIS \cite{24}. Among these variants, hybrid RIS has emerged as a promising architecture that comprises both active and passive elements, thereby enabling simultaneous signal amplification and reflection while maintaining moderate hardware complexity as well as power consumption. \cite{25} considered a hybrid RIS-aided communication network to jointly optimize transceiver beamforming at users and base station together with RIS reflection beamforming, demonstrating significant rate improvements over fully-passive RIS. The hybrid RIS element allocation problem for maximizing the ergodic capacity was elaborated in \cite{26}. Their results revealed that flexible assignment of active and passive elements could yield additional performance gains. \cite{Chen2024Physical} considered the deployment of hybrid RIS between the transmitter and legitimate receiver to boost secrecy capacity. Reference \cite{hardware1,hardware2,hardware3} accounted for hardware impairments (HWIs) at both phase-shifting elements and radio-frequency chains. Based on dedicated signal models, alternative optimization approaches were developed to maximize the achievable rate. Rigorous theoretical analyses further unveiled that different types of HWIs exhibit distinct rate-limiting behaviors. However, integrating hybrid RIS with digital AirComp induces a sophisticated tradeoff among channel-induced misalignment error, power amplification noise, and quantization distortion. To jointly accommodate these effects, our work formulates an optimization problem that targets E2E inference accuracy, which fundamentally differs from existing RIS-related studies that focus on AirComp error \cite{20,21,22,23,24} or communication rate \cite{25,26,Chen2024Physical,hardware1,hardware2,hardware3}. Besides, we develop a dedicated reflection beamforming approach to efficiently update both the amplitudes and phase shifts of the hybrid RIS, achieving significantly lower computational complexity than the widely adopted semidefinite relaxation framework \cite{21}.
    
    Table \ref{tab:comparison} explicitly compares this paper with related works.
    
    \begin{table*}[t]\footnotesize
    	\centering
    	\caption{Comparisons With Related Works}\label{tab:comparison}
    	\vspace{2.5 mm}
    	\begin{threeparttable}
    		\begin{tabular}{m{1.3cm}<{\centering}| m{1.3cm}<{\centering}| m{1.3cm}<{\centering}| m{1.5cm}<{\centering}| m{2.4cm}<{\centering}| m{2.8cm}<{\centering}| m{2.04cm}<{\centering}| m{1.5cm}<{\centering}}
    			\Xhline{1\arrayrulewidth}
    			\textbf{Reference} & \textbf{Edge AI Inference} &\textbf{AirComp Setup} &\textbf{Quantization Setup}& \textbf{RIS Setup} & \textbf{Theoretical Inference Performance Analysis}&\textbf{Quantization Bit Allocation}&\textbf{Beamforming Design}\\
    			\hline
    			\cite{8,9}  & \checkmark & Analog & \ding{55}& \ding{55} &\checkmark & \ding{55} & \ding{55}\\
    			\hline
    			\cite{10,12}   & \checkmark  & Analog & \ding{55}& \ding{55} &\checkmark & \ding{55} & \checkmark\\
    			\hline
    			\cite{11}   & \checkmark  & Analog & \ding{55}& \ding{55} &\ding{55} & \ding{55} & \checkmark\\
    			\hline
    			\cite{13,14,15}   & \ding{55}  & Digital & Scalar& \ding{55} &\ding{55} & \ding{55} & \ding{55}\\
    			\hline
    			\cite{16}   & \ding{55}  & Digital & Scalar& \ding{55} &\ding{55} & \checkmark & \ding{55}\\
    			\hline
    			\cite{17}   & \ding{55}  & Digital & Vector& \ding{55} &\ding{55} & \ding{55} & \ding{55}\\
    			\hline
    			\cite{18}   & \checkmark  & \ding{55} & \ding{55}& \ding{55} &\checkmark & \ding{55} & \ding{55}\\
    			\hline
    			\cite{19}   & \ding{55}  & \ding{55} & Vector& \ding{55} &\ding{55} & \checkmark & \ding{55}\\
    			\hline
    			\cite{20,21}   & \ding{55}  & Analog & \ding{55}& Passive RIS &\ding{55} & \ding{55} & \checkmark\\
    			\hline
    			\cite{22}   & \ding{55}  & Analog & \ding{55}& Active RIS &\ding{55} & \ding{55} & \checkmark\\
    			\hline
    			\cite{23}   & \ding{55}  & Analog & \ding{55}& STAR RIS &\ding{55} & \ding{55} & \checkmark\\
    			\hline
    			\cite{24}   & \ding{55}  & Analog & \ding{55}& Multi-functional RIS &\ding{55} & \ding{55} & \checkmark\\
    			\hline
    			\cite{25,26}   & \ding{55}  & \ding{55} & \ding{55}& Hybrid RIS &\ding{55} & \ding{55} & \checkmark\\
    			\hline
    			This paper  & \checkmark  & Digital & Vector &Hybrid RIS & \checkmark & \checkmark & \checkmark\\
    			\hline
    			\Xhline{1\arrayrulewidth}  
    		\end{tabular}
    	\end{threeparttable}
    	%\vspace{-0.25in}
    \end{table*}
    
    \subsection{Contributions}
    
    To empower communication-efficient feature aggregation for edge AI inference, this work proposes a Hybrid RIS-aided Digital AirComp (HRD-AirComp) scheme that is compatible with both existing and future mobile systems. We consider a scenario where multiple agents upload local features to an EN to collaboratively perform the inference task, with a hybrid RIS deployed to assist the over-the-air feature aggregation process. The proposed scheme exploits vector quantization to map each feature block comprising multiple local feature elements into a quantization codeword, whose index is subsequently employed to select a digitally modulated symbol sequence for transmission. By utilizing shared quantization and modulation codebooks across all agents, the waveform superposition inherent in AirComp can be leveraged to aggregate the quantization codewords at the EN, thereby facilitating global feature reconstruction. Furthermore, HRD-AirComp embraces a task-oriented design principle that identifies the importance levels of different feature elements and jointly optimizes quantization bit allocation as well as active-passive beamforming. This enables the preservation of critical features during transmission and ultimately improves inference accuracy. Our main contributions are summarized below.
    
    \begin{enumerate}
		\item To capture the unequal importance of local feature elements in edge inference, the proposed HRD-AirComp scheme partitions each high-dimensional feature vector into multiple blocks, which are quantized individually using differentiated codebooks with non-uniform bit allocation. The digital modulation codebook is then designed with a one-to-one mapping from the quantization codebook, enabling the EN to reconstruct each aggregated quantized feature block from the received superimposed signals across all agents. We present a performance analysis of the edge inference system exploiting HRD-AirComp-based feature aggregation. Leveraging the widely adopted entropy of posteriors metric, we derive a rigorous and tractable surrogate function for inference accuracy. This surrogate not only provides a closed-form characterization of the impact of distortions introduced by feature quantization and over-the-air aggregation, but also reveals an explicit feature importance indicator that can guide subsequent system optimization. 
		
		\item Building upon the theoretical analysis, we formulate an optimization problem that maximizes the surrogate of inference accuracy by jointly optimizing quantization bit allocation, agent transmission coefficients, EN receiving beamforming, and hybrid RIS reflection beamforming. The intricate coupling among these variables, together with the non-convex unit-modulus and amplification power constraints of the hybrid RIS, makes the problem highly challenging to solve. To this end, we develop an efficient algorithm that optimizes the four sets of variables in an alternating manner. Specifically, the bit allocation is determined using the successive convex approximation technique, while the transceiver designs at agents and EN are reformulated into convex optimization issues via a series of tailored transformations. For the hybrid RIS reflection beamforming, we propose a low-complexity iterative method that enables the closed-form update of both active and passive reflection coefficients. 
		
		\item We conduct extensive simulations to evaluate the performance of HRD-AirComp in inference tasks including both linear classification and multi-view object recognition. Experimental results demonstrate the convergence of the proposed optimization algorithm, as well as its superior implementation efficiency. The results also confirm the consistency between the derived surrogate function and the numerically computed inference accuracy, validating the effectiveness of our theoretical performance analysis. Besides, by harnessing the performance gains brought by adaptive feature quantization and hybrid RIS deployment, HRD-AirComp achieves 12.23\% to 24.67\% improvement in inference accuracy over state-of-the-art digital AirComp baselines, with performance close to that of the idealized case, which assumes perfect feature aggregation. 
	\end{enumerate}
    
    \textit{Notations:} Superscripts $\text{H}$, $\text{T}$, and $*$  denote the conjugate transpose, transpose, and conjugate, respectively. ${{\left[ \mathbf{X} \right]}_{i,j}}$ and ${{\left[ \mathbf{X} \right]}_{:,j}}$ are the $i$-th row and $j$-th column entry, and the $j$-th column of matrix $\mathbf{X}$. For a vector $\mathbf{a}$, $\left\| \mathbf{a} \right\|$, ${{\left[ \mathbf{a} \right]}_{i}}$, and $\text{diag}\left( \mathbf{a} \right)$ indicate its norm, the $i$-th element, and the matrix with diagonal elements composed by $\mathbf{a}$. ${{\mathbb{C}}^{M\times N}}$ and ${{\mathbb{R}}^{M\times N}}$ specify the space of $M\times N$ complex and real matrices. $\mathcal{C}\mathcal{N}\left( \bm{\mu} ,\mathbf{C} \right)$ and $\mathcal{N}\left( \bm{\mu} ,\mathbf{C} \right)$ represent the complex and real Gaussian distribution with centroid $\bm{\mu} $ and covariance $\mathbf{C}$. ${{\mathbf{1}}_{M\times N}}$ and ${{\mathbf{I}}_{M}}$ are $M\times N$ all-one matrix and $M\times M$ identity matrix, respectively. 
    
	\section{System Model} \label{sec:model}

\begin{figure}[t]
		\begin{center}
			\centerline{\includegraphics[width=6.5cm]{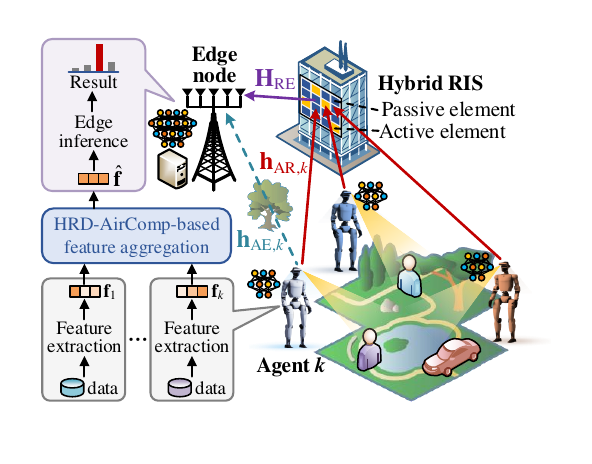}}
		\end{center}
		\vspace{-4mm}		
		\caption{Illustration of the edge inference system with a hybrid RIS.} 
		\label{fig:1}
	\end{figure}

	Consider an edge inference system designed for intelligent environmental perception, as illustrated in Fig. \ref{fig:1}. There are $K$ agents (whose set is denoted by $\mathcal{K}=\left\{ 1,\ldots ,k,\ldots ,K \right\}$) acquiring sensory data from distributed viewpoints, which are subsequently processed at an EN to generate inference results. During each perception round, each agent $k$ inputs its sensory data (e.g., images or 3D point clouds) into a light-weight AI model to extract local features. All local features are then transmitted wirelessly to the EN, with a hybrid RIS deployed to assist the feature uploading process. After receiving the superimposed signals from all agents, the EN reconstructs global features and performs edge inference\footnote{Without loss of generality, we focus on a single edge inference process, during which the channel state can be regarded as quasi-static due to the low-latency nature of AirComp \cite{8}. For continuous inference tasks spanning multiple time slots, the proposed scheme can be readily extended by repeatedly performing feature aggregation and re-optimizing the beamformers whenever the environment undergone notable changes.}. Detailed models are demonstrated in the following subsections. 
	
	\subsection{Inference Model}
	
	For each agent $k$, the features extracted by its local AI model are signified by a $W$-dimensional vector ${{\mathbf{f}}_{k}}\in {\mathbb{R}^{W\times 1}}$. Afterwards, all feature vectors $\left\{ {{\mathbf{f}}_{k}} \right\}_{k=1}^{K}$ undergo an aggregation process to yield a global feature vector $\mathbf{f}$ at the EN. Consider the popular average aggregation function, we have $\mathbf{f}=\frac{1}{K}\sum\nolimits_{k\in \mathcal{K}}{{{\mathbf{f}}_{k}}}$ in the ideal case. The EN then feeds $\mathbf{f}$ into its edge AI model to obtain the perception results (e.g., the label of the perceived target). 
	
	To address the communication bottleneck caused by high-dimensional feature uploading, the proposed edge inference system employs HRD-AirComp to enable efficient feature aggregation, while being compatible with the widely adopted digital communication paradigm. The specific operations of HRD-AirComp will be delineated in Section III, herein we define some required performance metrics. 
	
	Due to the feature quantization and channel distortion inherent in HRD-AirComp, the EN reconstructs a global feature vector $\mathbf{\hat{f}}$ that deviates from $\mathbf{f}$. The feature aggregation error is defined as $\mathbf{e}=\mathbf{\hat{f}}-\mathbf{f}$. Additionally, considering that a broad class of perception tasks can be formulated as classification problems, the inference accuracy at the EN is evaluated using the entropy of posteriors \cite{9}, which is written as
	\begin{align}
		H={{\mathbb{E}}_{{\mathbf{\hat{f}}}}}\left[ -\sum\limits_{l=1}^{L}{\Pr \left( l|\mathbf{\hat{f}} \right)\ln \Pr \left( l|\mathbf{\hat{f}} \right)} \right], \label{eq:1}
	\end{align}
	where $L$ indicates the number of classification classes, $\Pr ( l|\mathbf{\hat{f}} )$ represents the probability that the inference result corresponds to the $l$-th class (similar to the output of a softmax classifier in DNN). Accordingly, $H$ quantifies the uncertainty associated with the inference result given the reconstructed global features $\mathbf{\hat{f}}$. 
	
	\subsection{Communication Model}
	
	The EN and each agent are equipped with $M$ antennas and a single antenna, respectively. The hybrid RIS incorporates $N$ reflecting elements with ${{N}_{\text{a}}}$ active and ${{N}_{\text{p}}}$ passive elements, whose sets are denoted by ${{\mathcal{N}}_{\text{a}}}$ and ${{\mathcal{N}}_{\text{p}}}$, respectively. Let $\mathbf{\Phi }=\text{diag}\left( {{\phi }_{1}},\ldots ,{{\phi }_{N}} \right)\in {{\mathbb{C}}^{N\times N}}$ be the reflection matrix of the hybrid RIS, where ${{\phi }_{n}}$ represents the reflection coefficient of element $n$. Since both the amplitudes and phases of the active elements can be adjusted, whereas the passive elements are only capable of controlling the phase shifts, we have
	\begin{align}
		{{\phi }_{n}}=
		\left\{\begin{array}{cl}
			&\!\!\!\!\!\!\!\!\!\left| {{\phi }_{n}} \right|{{\text{e}}^{j{{\alpha }_{n}}}},\ n\in {{\mathcal{N}}_{\text{a}}},\\
			&\!\!\!\!\!\!\!\!\!{{\text{e}}^{j{{\alpha }_{n}}}},\qquad\;n\in {{\mathcal{N}}_{\text{p}}}, \\
		\end{array}\right.  \label{eq:2}
	\end{align}
    where ${{\alpha }_{n}}\in \left[ 0,2\pi  \right)$ denotes the phase shift of element $n$. To facilitate subsequent expressions, we decompose $\mathbf{\Phi }$ into 
    \begin{align}
       \mathbf{\Phi }={\text{diag}\left( \mathbf{1}_{N\times 1}^{\text{a}} \right)\mathbf{\Phi }}+{\text{diag}\left( \mathbf{1}_{N\times 1}^{\text{p}} \right)\mathbf{\Phi }}={{\mathbf{\Phi }}_{\text{a}}}+{{\mathbf{\Phi }}_{\text{p}}}, \label{eq:3}
    \end{align}
    where $\mathbf{1}_{N\times 1}^{\text{a}}$ and $\mathbf{1}_{N\times 1}^{\text{p}}$ are two $N$-dimensional binary vectors, and the indices of their non-zero elements are determined by ${{\mathcal{N}}_{\text{a}}}$ and ${{\mathcal{N}}_{\text{p}}}$, respectively. 
    
    Each agent $k$ modulates ${{\mathbf{f}}_{k}}$ into $T$ transmission signals, denoted as $\left\{ {{\mathbf{s}}_{k,t}} \right\}_{t=1}^{T}\in {{\mathbb{C}}^{J\times 1}}$, and $J$ is the length of each signal. Let ${{\mathbf{h}}_{\text{AE},k}}\in {{\mathbb{C}}^{M\times 1}}$, ${{\mathbf{h}}_{\text{AR},k}}\in {{\mathbb{C}}^{N\times 1}}$, and ${{\mathbf{H}}_{\text{RE}}}\in {{\mathbb{C}}^{M\times N}}$ represent the channel from agent $k$ to EN, that from agent $k$ to hybrid RIS, and that from hybrid RIS to EN, respectively. After all agents simultaneously transmit signals $\left\{ {{\mathbf{s}}_{k,t}} \right\}_{k=1}^{K}$, the received signal at the EN is written as
    \begin{align}
    	{{\mathbf{y}}_{t}}=\left[ \sum\limits_{k\in \mathcal{K}}{{{\nu }_{k}}{{\mathbf{s}}_{k,t}}\mathbf{h}_{k}^{\text{T}}}+{{\mathbf{Z}}_{\text{R},t}}{{\mathbf{\Phi }}_{\text{a}}}\mathbf{H}_{\text{RE}}^{\text{T}}+{{\mathbf{Z}}_{\text{E},t}} \right]\mathbf{b}\in {{\mathbb{C}}^{J\times 1}}, \label{eq:4}
    \end{align}
    where ${{\nu }_{k}}$ signifies the transmission coefficient of agent $k$. ${{\mathbf{h}}_{k}}={{\mathbf{h}}_{\text{AE},k}}+{{\mathbf{H}}_{\text{RE}}}\mathbf{\Phi }{{\mathbf{h}}_{\text{AR},k}}$ indicates the effective channel vector between agent $k$ and the EN. ${{\mathbf{Z}}_{\text{R},t}}\in {{\mathbb{C}}^{J\times N}}$ characterizes the thermal noise incurred by the active elements of the hybrid RIS, which is modeled as ${{\left[ {{\mathbf{Z}}_{\text{R},t}} \right]}_{j,n\in {{\mathcal{N}}_{\text{a}}}}}\sim \mathcal{C}\mathcal{N}\left( 0,\sigma _{\text{R}}^{2} \right)$ and ${{\left[ {{\mathbf{Z}}_{\text{R},t}} \right]}_{j,n\in {{\mathcal{N}}_{\text{p}}}}}=0$ with $\sigma _{\text{R}}^{2}$ being the noise variance\footnote{The noise introduced at the hybrid RIS is multiplied by amplification coefficients ${{\mathbf{\Phi }}_{\text{a}}}$ and channel $\mathbf{H}_{\text{RE}}$, thus its impact on the EN-side received signal depends on the reflection beamforming \cite{25,33}.}. ${{\mathbf{Z}}_{\text{E},t}}\in {{\mathbb{C}}^{J\times M}}$ indicates the antenna noise at the EN, in which each element obeys $\mathcal{C}\mathcal{N}\left( 0,\sigma _{\text{E}}^{2} \right)$ with variance $\sigma _{\text{E}}^{2}$. $\mathbf{b}\in {{\mathbb{C}}^{M\times 1}}$ denotes the receiving beamformer. 
    
    \section{HRD-AirComp-Based Feature Aggregation Scheme}
    
    \begin{figure}[t] \centering
    	\begin{center}
    		\centerline{\includegraphics[width=8.5cm]{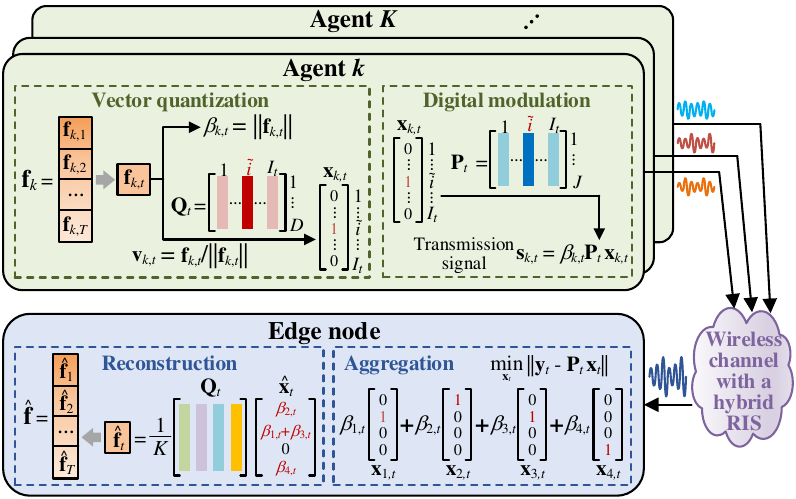}}
    	\end{center}
    	\vspace{-4mm}		
    	\caption{The diagram of HRD-AirComp scheme.}
    	\label{fig:2}     
    \end{figure}

    The operations of the proposed HRD-AirComp scheme are illustrated in Fig. \ref{fig:2} and elaborated in this section. The key innovation lies in the seamless integration of vector quantization and digital modulation into AirComp. Rather than directly aggregating analog values, the scheme leverages waveform superposition to support the aggregation of quantization codewords. Furthermore, we emphasize that HRD-AirComp serves as a general approach for over-the-air data computation in digital communication systems, with this work focusing on its application to feature aggregation in edge inference. 
    
    \subsection{Vector Quantization}
    
    At each agent $k$, the $W$-dimensional feature vector ${{\mathbf{f}}_{k}}$ is divided into $T$ blocks\footnote{As each block is quantized and then modulated into a transmission signal, they have the same number $T$. In particular, introducing vector quantization into digital AirComp not only significantly reduces the communication overhead (by a factor of $D$ compared with scalar quantization), but also brings advantages in exploiting the feature’s geometric structure \cite{19,28}.} of length $D$, represented by $\mathbf{f}_{k}^{\text{T}}=[ \mathbf{f}_{k,1}^{\text{T}},\ldots ,\mathbf{f}_{k,T}^{\text{T}} ]$ with ${{\mathbf{f}}_{k,t}}\in \mathbb{R}^{D\times 1}$ indicating the $t$-th block, and we have $W=TD$. Each block, say ${{\mathbf{f}}_{k,t}}$, is further decomposed into its norm ${{\beta }_{k,t}}=\left\| {{\mathbf{f}}_{k,t}} \right\|$ and normalization ${{\mathbf{v}}_{k,t}}={{{\mathbf{f}}_{k,t}}}/{\left\| {{\mathbf{f}}_{k,t}} \right\|}$. Owing to the scalar ${{\beta }_{k,t}}$ can be conveyed through the amplitude of the transmission signal, we concentrate on the quantization of vector ${{\mathbf{v}}_{k,t}}$ in the sequel. 
    
    Since ${{\mathbf{v}}_{k,t}}$ is a unit vector uniformly distributed over the hypersphere in ${{\mathbb{R}}^{D\times 1}}$, it defines a one-dimensional subspace that resides on the Grassmannian manifold $\mathcal{G}\left( 1,D \right)$ \cite{27}. Consequently, ${{\mathbf{v}}_{k,t}}$ can be quantized using a Grassmannian codebook ${{\mathbf{Q}}_{t}}\in {{\mathbb{R}}^{D\times {{I}_{t}}}}$, with ${{I}_{t}}$ being the codebook size. Given that different feature dimensions contribute unequally to edge inference, the bit allocation across quantization codebooks corresponding to different blocks is non-uniform. Let $B$ denote the total number of available bits, and ${{B}_{t}}$ is the number of bits assigned to codebook ${{\mathbf{Q}}_{t}}$, such that ${{I}_{t}}={{2}^{{{B}_{t}}}}$ and $\sum\nolimits_{t=1}^{T}{{{B}_{t}}}=B$. Note that $\left\{ {{B}_{t}} \right\}_{t=1}^{T}$ are optimization variables and will be designed to enhance inference accuracy in Section V. With given ${{B}_{t}}$, the corresponding codebook ${{\mathbf{Q}}_{t}}$ can be optimized using the Lloyd algorithm in offline \cite{28}, which iteratively updates clusters on $\mathcal{G}\left( 1,D \right)$. The resulting cluster centroids form the quantization codewords of ${{\mathbf{Q}}_{t}}$. 
    
    Subsequently, ${{\mathbf{Q}}_{t}}$ is shared by all agents to quantize $\left\{ {{\mathbf{v}}_{k,t}} \right\}_{k=1}^{K}$. For each agent $k$, let ${{\mathbf{v}}_{k,t}}$ be quantized into the $\tilde{i}$-th codeword in ${{\mathbf{Q}}_{t}}$ with the minimum distance, i.e., 
    \begin{align}
    	\tilde{i}=\arg \underset{i\in \left\{ 1,\ldots ,{{I}_{t}} \right\}}{\mathop{\min }}\,\left\| {{\mathbf{v}}_{k,t}}-{{\left[ {{\mathbf{Q}}_{t}} \right]}_{:,i}} \right\|. \label{eq:5}
    \end{align}
    We then define a ${{I}_{t}}$ -dimensional binary vector ${{\mathbf{x}}_{k,t}}$, in which ${{\left[ {{\mathbf{x}}_{k,t}} \right]}_{{\tilde{i}}}}=1$, and the other ${{I}_{t}}-1$ elements are zero. As such, the quantized version of ${{\mathbf{v}}_{k,t}}$ is given by ${{\mathbf{Q}}_{t}}{{\mathbf{x}}_{k,t}}$. 
    
    \subsection{Digital Modulation}
    
    After obtaining the quantization codeword, each agent modulates it into a transmission sequence. To this end, we introduce a modulation codebook ${{\mathbf{P}}_{t}}\in {{\mathbb{C}}^{J\times {{I}_{t}}}}$ shared by all agents, where each column represents a sequence of length $J$, and a one-to-one mapping exists between the ${{I}_{t}}$ columns of ${{\mathbf{P}}_{t}}$ and those of ${{\mathbf{Q}}_{t}}$. To be specific, if block ${{\mathbf{v}}_{k,t}}$ is quantized into ${{\left[ {{\mathbf{Q}}_{t}} \right]}_{:,\tilde{i}}}$, then it is modulated as sequence ${{\left[ {{\mathbf{P}}_{t}} \right]}_{:,\tilde{i}}}$. Notably, the sequences in ${{\mathbf{P}}_{t}}$ can be generated based on standard digital modulation constellations, ensuring that the proposed HRD-AirComp scheme is highly compatible with current communication systems \cite{29}. By multiplying the selected sequence with block norm ${{\beta }_{k,t}}$, the transmission signal of each agent $k$ can be expressed as ${{\mathbf{s}}_{k,t}}={{\beta }_{k,t}}{{\mathbf{P}}_{t}}{{\mathbf{x}}_{k,t}}$. 
    
    \subsection{Feature Aggregation and Reconstruction}
    
    Given $\left\{ {{\mathbf{s}}_{k,t}} \right\}_{k=1}^{K}$, the EN aims to compute ${{\mathbf{s}}_{t}}=\sum\nolimits_{k\in \mathcal{K}}{{{\mathbf{s}}_{k,t}}}={{\mathbf{P}}_{t}}\sum\nolimits_{k\in \mathcal{K}}{{{\beta }_{k,t}}{{\mathbf{x}}_{k,t}}}$ via AirComp. We then define ${{\mathbf{x}}_{t}}=\sum\nolimits_{k\in \mathcal{K}}{{{\beta }_{k,t}}{{\mathbf{x}}_{k,t}}}$, and it can be seen that its $i$-th element ${{\left[ {{\mathbf{x}}_{t}} \right]}_{i}}$ equals $\sum\nolimits_{k\in {{\mathcal{K}}_{i}}}{{{\beta }_{k,t}}}$, with ${{\mathcal{K}}_{i}}$ denoting the set of agents which quantize their block normalizations ${{\left\{ {{\mathbf{v}}_{k,t}} \right\}}_{k\in {{\mathcal{K}}_{i}}}}$ into codeword ${{\left[ {{\mathbf{Q}}_{t}} \right]}_{:,i}}$. In practice, the EN first observes the received signal ${{\mathbf{y}}_{t}}$ in (\ref{eq:4}), which serves as an estimation of ${{\mathbf{s}}_{t}}$, then detects ${{\mathbf{x}}_{t}}$ using the known modulation codebook ${{\mathbf{P}}_{t}}$. The detection result of ${{\mathbf{x}}_{t}}$ at the EN is expressed as ${{\mathbf{\hat{x}}}_{t}}$. To ensure that ${{\mathbf{\hat{x}}}_{t}}$ closely approximates ${{\mathbf{x}}_{t}}$, it is essential to jointly design the agent transmission coefficients, EN receiving beamforming, and hybrid RIS reflection beamforming. A received signal processing algorithm should be also considered to detect ${{\mathbf{x}}_{t}}$, which will be elaborated in Section V. 
    
    Thereafter, the average of the $t$-th feature blocks of all agents, i.e., ${{\mathbf{f}}_{t}}=\frac{1}{K}\sum\nolimits_{k\in \mathcal{K}}{{{\mathbf{f}}_{k,t}}}$, can be estimated by ${{\mathbf{\hat{f}}}_{t}}=\frac{1}{K}{{\mathbf{Q}}_{t}}{{\mathbf{\hat{x}}}_{t}}$. By cascading all the $T$ aggregated blocks, the EN reconstructs the global feature vector as follows
	\begin{align}
		\mathbf{\hat{f}}\!\!=\!\!{{\left[ \mathbf{\hat{f}}_{1}^{\text{T}},\mathbf{\hat{f}}_{2}^{\text{T}},\ldots ,\mathbf{\hat{f}}_{T}^{\text{T}} \right]}^{\text{T}}}\!\!\!\!=\!\!\frac{1}{K}\!{{\left[\! {{\left( {{\mathbf{Q}}_{1}}{{{\mathbf{\hat{x}}}}_{1}} \right)\!}^{\text{T}}}\!,\!{{\left( {{\mathbf{Q}}_{2}}{{{\mathbf{\hat{x}}}}_{2}} \right)\!}^{\text{T}}}\!,\!\ldots,\!{{\left( {{\mathbf{Q}}_{T}}{{{\mathbf{\hat{x}}}}_{T}} \right)\!}^{\text{T}}} \right]\!}^{\text{T}}}\!\!. \label{eq:6}
	\end{align}
	
	\textit{Remark 1 (Implementation Complexity of HRD-AirComp):} As the Grassmannian codebooks are optimized offline and pre-stored at all agents, the implementation complexity only arises from searching the nearest quantization codeword for each feature block. The resulting complexity is $\mathcal{O}( \sum_{t=1}^{T} I_t D )$, which scales linearly with feature dimension and codebook size, indicating a low computational overhead at the agents.
	
	\section{Theoretical Analysis on Inference Accuracy With HRD-AirComp} 
	
	In this section, we theoretically analyze the inference performance of the proposed system exploiting HRD-AirComp-based feature aggregation. Given the intractable form of the inference accuracy metric $H$, the analysis proceeds in two main steps: characterizing the distribution of reconstructed global features, and designing reasonable surrogate function for $H$. The resulting insights offer theoretical guidance for the system optimization in Section V. 
	
	We commence by introducing some commonly adopted assumptions in inference performance analysis.
	
	\textit{Assumption 1:} The error-free version of the global feature vector, $\mathbf{f}$, follows a Gaussian mixture distribution as below\cite{9}
	\begin{align}
		\mathbf{f}\sim \frac{1}{L}\sum\limits_{l=1}^{L}{\mathcal{N}\left( {{\bm{\mu} }_{l}},\mathbf{C} \right)},  \label{eq:7}
	\end{align}
	where ${\bm{\mu}_{l}}={{\left[ {{\mu }_{l,1}},\ldots ,{{\mu }_{l,W}} \right]}^{\text{T}}}\in {{\mathbb{R}}^{W\times 1}}$ denotes the centroid of the $l$-th class, $\mathbf{C}=\text{diag}\left( {{c}_{1}},\ldots ,{{c}_{W}} \right)\in {{\mathbb{R}}^{W\times W}}$ is the diagonal covariance matrix.
	
	\textit{Assumption 2:} The EN detects ${{\mathbf{x}}_{t}}$ from a noised observation ${{\mathbf{y}}_{t}}$, and the detection mean square error (MSE) is determined by the computational MSE between ${{\mathbf{y}}_{t}}$ and the desired result ${{\mathbf{s}}_{t}}$, namely $\mathbb{E}[ {{\left\| {{{\mathbf{\hat{x}}}}_{t}}-{{\mathbf{x}}_{t}} \right\|}^{2}} ]=\eta \mathbb{E}[ {{\left\| {{\mathbf{y}}_{t}}-{{\mathbf{s}}_{t}} \right\|}^{2}} ]$ with $\eta $ being a constant \cite{17}. Notice that this assumption is commonly adopted to characterize the estimation error of compressive sensing algorithms, as considered in \cite{17} and Section V-D in this paper. It is independent of the quantization procedure and only pertains to the quality of sparse signal recovery.
	
	\textit{Assumption 3:} Since all agents extract local features from a shared data source, the transmission signals, modulated from their quantized features, are inherently correlated. Such correlations can be characterized by a matrix $\mathbf{U}\in {{\mathbb{R}}^{K\times K}}$, whose $\left( k,{k}' \right)$-th entry is given by \cite{12}
	\begin{align}
		{{u}_{k,{k}'}}=\frac{1}{J}\mathbb{E}\left[ {{\left( {{\mathbf{P}}_{t}}{{\mathbf{x}}_{k,t}} \right)}^{\text{H}}}{{\mathbf{P}}_{t}}{{\mathbf{x}}_{{k}',t}} \right],\forall t. \label{eq:8}
	\end{align}
    Hence, ${{u}_{k,{k}'}}$ indicates the correlation between the selected sequences of agent $k$ and ${k}'$, while $J$ is used for normalizing ${{u}_{k,{k}'}}\in \left[ 0,1 \right]$. 
    
    \textit{Assumption 4:} The block norms of all agents are upper bounded by $\mathbb{E}\left[ {{\beta }_{k,t}} \right]\le \beta ,\forall k,t$ \cite{19}. 
    
    \subsection{Distribution of Reconstructed Global Features}
    
    To calculate the probabilities $\Pr ( l|\mathbf{\hat{f}} ),\forall l$ in the entropy of posteriors $H$ in (\ref{eq:1}), we need to derive the distribution of the reconstructed global feature vector $\mathbf{\hat{f}}$. Based on the operations of HRD-AirComp, we first characterize the feature aggregation error $\mathbf{e}=\mathbf{\hat{f}}-\mathbf{f}$ as follows. 
    
    \textit{Theorem 1:} Under Assumption 2-4, the aggregation MSE of the $t$-th block, $\mathbb{E}[ {{\| {{{\mathbf{\hat{f}}}}_{t}}-{{\mathbf{f}}_{t}} \|}^{2}} ]$, has an upper bound given by 
    \begin{align}
    	&\!\!\mathbb{E}[ {{\| {{{\mathbf{\hat{f}}}}_{t}}-{{\mathbf{f}}_{t}} \|}^{2}} ] \nonumber\\
    	&\ \le\underbrace{\frac{{{2}^{{{B}_{t}}+1}}\eta {{\beta }^{2}}J}{{{K}^{2}}}\sum\limits_{k\in \mathcal{K}}{\sum\limits_{{k}'\in \mathcal{K}}\!\!{{{u}_{k,{k}'}}{{\left( {{\nu }_{k}}\mathbf{h}_{k}^{\text{T}}\mathbf{b}\!-\!1 \right)}^{*}}\!\left( {{\nu }_{{{k}'}}}\mathbf{h}_{{{k}'}}^{\text{T}}\mathbf{b}\!-\!1 \right)}}}_{\text{Misalignment error}}\nonumber\\
    	&\ +\underbrace{\frac{{{2}^{{{B}_{t}}+1}}\eta }{{{K}^{2}}}\!\left[ \sigma _{\text{R}}^{2}{{\left\| {{\mathbf{\Phi }}_{\text{a}}}\mathbf{H}_{\text{RE}}^{\text{T}}\mathbf{b} \right\|}^{2}}\!\!+\!\sigma _{\text{E}}^{2}{{\left\| \mathbf{b} \right\|}^{2}} \right]}_{\text{Channel noise}}+\!\!\!\!\!\underbrace{\frac{{{\beta }^{2}}}{K}{{2}^{1\!-\!\frac{2{{B}_{t}}}{D-1}}}}_{\text{Quantization distortion}}\!\!\!\!\!\!\triangleq {{\varepsilon }_{t}}, \label{eq:9}
    \end{align}

    \textit{Proof:} Please refer to Appendix A. $\qquad\qquad\qquad\qquad\ \ \blacksquare$
    
    \textit{Remark 2:} Theorem 1 provides critical insights into the aggregation error of the proposed HRD-AirComp scheme. \textit{First}, the overall error comprises three components: misalignment error, channel noise, and quantization distortion. The first two components are common in conventional analog AirComp \cite{20}, whereas the third arises from the distortion introduced by vector quantization and digital modulation. \textit{Second}, traditional AirComp typically assumes independent transmission signals across agents \cite{23}, which is not hold in edge inference systems. To account for the feature correlations, the misalignment error in HRD-AirComp incorporates inter-agent coupling, making it more intricate than in conventional AirComp. \textit{Third}, the deployment of hybrid RIS can enhance the agent-EN channels, but also incur additional channel noise. This calls for the judicious design of active and passive reflection coefficients at the hybrid RIS to fully exploit its performance gains. \textit{Fourth}, quantization bit allocation introduces a tradeoff, i.e., increasing ${{B}_{t}}$ reduces the quantization error for the $t$-th block, but also amplifies both the misalignment error and channel noise. Therefore, $\left\{ {{B}_{t}} \right\}_{t=1}^{T}$ should be carefully allocated to protect the important features. \textit{Fifth}, we can observe that ${{\varepsilon }_{t}}$ decreases with the increment of $K$, as more agents provide larger view and channel diversity to suppress aggregation error. The growth of number of antennas $M$ and RIS elements $N$ offers higher DoFs for adjusting $\mathbf{b}$ and $\{\mathbf{h}_{k}\}$, thereby reducing misalignment error.
    
    Afterwards, leveraging the central limit theorem \cite{30}, the feature aggregation error $\mathbf{e}$ can be approximated as a Gaussian distribution with zero mean and variance
    \begin{align}
    	{{\mathbf{C}}^{\mathbf{e}}}\!=\!\frac{1}{D}\text{diag}\left( \underbrace{\varepsilon _{1}^{2},\ldots ,\varepsilon _{1}^{2}}_{\text{Number of }D},\underbrace{\varepsilon _{2}^{2},\ldots ,\varepsilon _{2}^{2}}_{\text{Number of }D},\ldots ,\underbrace{\varepsilon _{T}^{2},\ldots ,\varepsilon _{T}^{2}}_{\text{Number of }D} \right)\!,\! \label{eq:10}
    \end{align}
    where $\varepsilon _{t},t\in\{1,\dots,T\}$ is defined in (\ref{eq:9})\footnote{Given that the $t$-th entry in $\mathbf{e}$ has zero mean, its variance coincides with the second-order moment, i.e., $\mathbb{E}[ {{\| {{{\mathbf{\hat{f}}}}_{t}}-{{\mathbf{f}}_{t}} \|}^{2}}]$, which is approximated by the upper bound $\varepsilon _{t}$ for worst-case analysis.}, and the $\left( w,w \right)$-th entry of ${{\mathbf{C}}^{\mathbf{e}}}$ is represented by $c_{w}^{\mathbf{e}}$. Then we have the following corollary to characterize the distribution of the reconstructed global features. 
    
    \textit{Corollary 1:} Based on Theorem 1 and Assumption 1, the reconstructed global feature vector $\mathbf{\hat{f}}$ follows a Gaussian mixture distribution
    \begin{align}
    	\mathbf{\hat{f}}\sim \frac{1}{L}\sum\limits_{l=1}^{L}{\mathcal{N}\left( {{{\hat{\bm{\mu} }}}_{l}},\mathbf{\hat{C}} \right)}, \label{eq:11}
    \end{align}
    where ${{\hat{\bm{\mu} }}_{l}}={{\bm{\mu} }_{l}}$, $\mathbf{\hat{C}}=\mathbf{C}+{{\mathbf{C}}^{\mathbf{e}}}$. 
    
    \textit{Proof:} According to the definition $\mathbf{\hat{f}}=\mathbf{f}+\mathbf{e}$, the $l$-th component of $\mathbf{\hat{f}}$ is expressed as the summation of independent Gaussian vectors, with mean ${{\bm{\mu} }_{l}}$ and variance $\mathbf{C}+{{\mathbf{C}}^{\mathbf{e}}}$. By averaging all the $L$ components, we can obtain (\ref{eq:11}). This completes the proof. $\qquad\qquad\qquad\qquad\qquad\qquad\qquad\qquad\blacksquare$
    
    The distribution of $\mathbf{\hat{f}}$ lays a foundation for further deducing the inference accuracy, as elaborated in the next subsection. 
    
    \subsection{Surrogate Function for Inference Accuracy}
    
    Given the inference accuracy metric, i.e., the entropy of posteriors $H$ in (\ref{eq:1}), we establish the following theorem to provide a rigorous lower bound.
    
    \textit{Theorem 2:} Based on Corollary 1, the entropy of posteriors $H$ can be lower bounded by
    \begin{align}
    	H\!\ge\! \frac{1}{L}\sum\limits_{l=1}^{L}{\ln \left[ 1\!+\!{{G}_{l,{l}'}} \right]}\!\approx\! \frac{1}{L}\sum\limits_{l=1}^{L}{\ln \left[ 1\!+\!\left( L\!-\!1 \right){{\text{e}}^{{-G}/{2}}} \right]},  \label{eq:12}
    \end{align}
    where ${{G}_{l,{l}'}}={{\left( {{{\hat{\bm{\mu} }}}_{l}}-{{{\hat{\bm{\mu} }}}_{{{l}'}}} \right)}^{\text{T}}}{{\mathbf{\hat{C}}}^{-1}}\left( {{{\hat{\bm{\mu} }}}_{l}}-{{{\hat{\bm{\mu} }}}_{{{l}'}}} \right)$, and $G=\frac{1}{L\left( L-1 \right)}\sum\nolimits_{l<{l}'\le L}{{{G}_{l,{l}'}}}$. 
    
    \textit{Proof:} Please refer to Appendix B. $\qquad\qquad\qquad\qquad\ \ \blacksquare$
    
    \textit{Remark 3:} Theorem 2 yields two key observations. \textit{First}, the lower bound of $H$ is determined by $\left\{ {{G}_{l,{l}'}} \right\}$ and $G$, both of which possess clear geometric interpretations. Specifically, ${{G}_{l,{l}'}}$ quantifies the distance between the centroids of class pair $\left( l,{l}' \right)$, normalized by the covariance. A larger ${{G}_{l,{l}'}}$ indicates that the two classes are more distinguishable in the feature space. $G$ represents the average of $\left\{ {{G}_{l,{l}'}} \right\}$ across all class pairs, thus capturing the overall class separability given the reconstructed global features. \textit{Second}, the lower bound of $H$ is shown to be monotonically decreasing with respect to (w.r.t.) $G$, implying that increasing $G$ leads to lower inference uncertainty and thus higher accuracy. As a consequence, $G$ serves as a tractable surrogate for the original metric $H$. 
    
    To this end, we rewrite $G$ as follows, which offers a task-oriented objective for system optimization. 
	\begin{align}
		\!\! G\!=\!\frac{1}{L\left( L\!-\!1 \right)}\sum\limits_{l<{l}'\le L}{\sum\limits_{w=1}^{W}{\frac{{{\left( {{\mu }_{l,w}}\!-\!{{\mu }_{{l}',w}} \right)}^{2}}}{{{c}_{w}}+c_{w}^{\mathbf{e}}}}}\!=\!\sum\limits_{w=1}^{W}\!{\frac{{{\rho }_{w}}}{{{c}_{w}}\!+\!c_{w}^{\mathbf{e}}}}. \label{eq:13}
	\end{align}
    Here, ${{{\rho }_{w}}=\sum\nolimits_{l<{l}'\le L}{{{\left( {{\mu }_{l,w}}-{{\mu }_{{l}',w}} \right)}^{2}}}}/{L\left( L-1 \right)}$ can be interpreted as an importance indicator for the $w$-th feature dimension. It is desirable to allocate more resources to suppress the variance $c_{w}^{\mathbf{e}}$ of dimensions with larger ${{\rho }_{w}}$, thereby enhancing the class separability $G$ and ultimately improving the edge inference accuracy. 
    
    \section{Joint Feature Quantization and Active-Passive Beamforming Design}
    
    Building on the surrogate for inference accuracy established in the preceding section, we now aim to maximize this surrogate by developing a joint optimization algorithm for feature quantization and active-passive beamforming design. 
    
    \subsection{ Problem Formulation and Transformation}
    
    With the objective of maximizing $G$ in (\ref{eq:13}), the problem for jointly optimizing quantization bit allocation $\left\{ {{B}_{t}} \right\}_{t=1}^{T}$, agent transmission coefficients $\left\{ {{\nu }_{k}} \right\}_{k=1}^{K}$, EN receiving beamforming $\mathbf{b}$, and hybrid RIS reflection beamforming $\mathbf{\Phi }$ can be formulated as
    \begin{subequations}
    	\begin{equation}
    		\begin{aligned}
    			\textbf{P1}: \underset{\left\{ {{B}_{t}} \right\}_{t=1}^{T},\left\{ {{\nu }_{k}} \right\}_{k=1}^{K},\mathbf{b},\mathbf{\Phi }}{\mathop{\max }}\,\sum\limits_{w=1}^{W}{\frac{{{\rho }_{w}}}{{{c}_{w}}+c_{w}^{\mathbf{e}}}}, \label{eq:14a}
    		\end{aligned}
    	\end{equation}
    	\vspace{-5mm}
    	\begin{align}
    		\mbox{s.t.}\ 
    		&\sum\nolimits_{t=1}^{T}{{{B}_{t}}}=B, \label{eq:14b}\\
    		&{{\left| {{\nu }_{k}} \right|}^{2}}\le \frac{{{P}_{\text{A}}}}{{{\beta }^{2}}J},\ \forall k, \label{eq:14c}\\
    		&{{\alpha }_{n}}\in \left[ 0,2\pi  \right),\ \forall n, \label{eq:14d}\\
    		&\left| {{\left[ \mathbf{\Phi } \right]}_{n,n}} \right|=1,\text{ }\forall n\in {{\mathcal{N}}_{\text{p}}}, \label{eq:14e}\\
    		&\text{tr}\big[ \mathbf{\Phi }_{\text{a}}^{\text{H}}\big( {{\beta }^{2}}J\sum\nolimits_{k\in \mathcal{K}}{\sum\nolimits_{{k}'\in \mathcal{K}}{{{u}_{k,{k}'}}\nu _{k}^{*}{{\nu }_{{{k}'}}}\mathbf{h}_{\text{AR},k}^{*}\mathbf{h}_{\text{AR},{k}'}^{\text{T}}}}\nonumber\\
    		&\qquad\qquad\qquad\quad\ \ +\sigma _{\text{R}}^{2}{{\mathbf{I}}_{N}} \big){{\mathbf{\Phi }}_{\text{a}}} \big]\le {{P}_{\text{R}}}, \label{eq:14f}
    	\end{align} 
    \end{subequations}
    where constraint (\ref{eq:14b}) limits the total available number of bits. (\ref{eq:14c}) indicates the transmission power constraint with ${{P}_{\text{A}}}$ being agent power budget. (\ref{eq:14d}) restricts the phase shift at the hybrid RIS. (\ref{eq:14e}) demonstrates that the passive elements should satisfy unit-modulus constraints. In (\ref{eq:14f}), the amplification power of the hybrid RIS is not higher than its power budget ${{P}_{\text{R}}}$. 
    
    To cope with the complex optimization objective in (\ref{eq:14a}) in the form of summation over multiple fractional functions, we introduce auxiliary variables $\left\{ {{\lambda }_{w}} \right\}_{w=1}^{W}$ with ${{\lambda }_{w}}=\frac{{{\rho }_{w}}}{{{c}_{w}}+c_{w}^{\mathbf{e}}},\text{ }\forall w$, and transform \textbf{P1} into
    \begin{subequations}
    	\begin{equation}
    		\begin{aligned}
    			\textbf{P2}: \underset{\left\{ {{B}_{t}} \right\}_{t=1}^{T},\left\{ {{\nu }_{k}} \right\}_{k=1}^{K},\mathbf{b},\mathbf{\Phi },\left\{ {{\lambda }_{w}} \right\}_{w=1}^{W}}{\mathop{\max }}\,\sum\limits_{w=1}^{W}{{{\lambda }_{w}}}, \label{eq:15a}
    		\end{aligned}
    	\end{equation}
    	\vspace{-5mm}
    	\begin{align}
    		\mbox{s.t.}\ 
    		&c_{w}^{\mathbf{e}}\le \frac{{{\rho }_{w}}}{{{\lambda }_{w}}}-{{c}_{w}},\ \forall w, \label{eq:15b}\\
    		&\text{(\ref{eq:14b})}\sim \text{(\ref{eq:14f})},\nonumber
    	\end{align} 
    \end{subequations}
    where the equality in constraint (\ref{eq:15b}) must hold at the optimal solution of \textbf{P1}, otherwise we can increase ${{\lambda }_{w}}$ to enlarge the objective value, thus the transformation does not impact the optimality. However, solving \textbf{P2} remains challenging due to the strong coupling between quantization bit allocation and high-dimensional beamforming variables, as well as the correlation among agents’ transmission signals. Additionally, the hybrid RIS imposes both unit-modulus constraint at its passive elements as well as amplification power constraint at active elements, making the reflection beamforming design more complicated than that in conventional RIS-aided AirComp. 
    
    To tackle these challenges, we decompose \textbf{P2} into four subproblems, which optimizes bit allocation $\left\{ {{B}_{t}} \right\}_{t=1}^{T}$, transmission coefficients $\left\{ {{\nu }_{k}} \right\}_{k=1}^{K}$, receiving beamforming $\mathbf{b}$, and reflection beamforming $\mathbf{\Phi }$ in an alternative fashion. For ease of discerning, the variables with bar, e.g., ${{\bar{\nu }}_{k}}$, means that they are obtained from the previous iteration. 
    
    \subsection{Quantization Bit Allocation Optimization}
     
    The subproblem for determining $\left\{ {{B}_{t}} \right\}_{t=1}^{T}$ with fixed $\left\{ {{\nu }_{k}} \right\}_{k=1}^{K},\mathbf{b},\mathbf{\Phi }$ is given by
	\begin{subequations}
		\begin{equation}
			\begin{aligned}
				\textbf{SP1}: \underset{\left\{ {{B}_{t}} \right\}_{t=1}^{T},\left\{ {{\lambda }_{w}} \right\}_{w=1}^{W}}{\mathop{\max }}\,\sum\limits_{w=1}^{W}{{{\lambda }_{w}}}, \label{eq:16a}
			\end{aligned}
		\end{equation}
		\vspace{-5mm}
		\begin{align}
			\mbox{s.t.}\ 
			&{{\psi }_{1}}{{2}^{{{B}_{t}}}}+{{\psi }_{2}}{{2}^{-\frac{2{{B}_{t}}}{D-1}}}\le \frac{{{\rho }_{w}}}{{{\lambda }_{w}}}-{{c}_{w}},\ \forall t\in \left\{ 1,\ldots ,T \right\},\nonumber\\
			&\qquad\qquad\quad\forall w\in \left\{ \left( t-1 \right)D+1,\ldots ,tD \right\}, \label{eq:16b}\\
			&\text{(\ref{eq:14b})},\nonumber
		\end{align} 
	\end{subequations}
	where ${{\psi }_{1}}=\frac{2\eta {{\beta }^{2}}J}{{{K}^{2}}D}\!\sum\limits_{k\in \mathcal{K}}{\sum\limits_{{k}'\in \mathcal{K}}\!\!{{{u}_{k,{k}'}}{{\left( {{{\bar{\nu }}}_{k}}\mathbf{\bar{h}}_{k}^{\text{T}}\mathbf{\bar{b}}\!-\!1 \right)}^{*}}\left( {{{\bar{\nu }}}_{{{k}'}}}\mathbf{\bar{h}}_{{{k}'}}^{\text{T}}\mathbf{\bar{b}}\!-\!1 \right)}}+\frac{2\eta }{{{K}^{2}}D}\left[ \sigma _{\text{R}}^{2}{{\left\| {{{\mathbf{\bar{\Phi }}}}_{\text{a}}}\mathbf{H}_{\text{RE}}^{\text{T}}\mathbf{\bar{b}} \right\|}^{2}}+\sigma _{\text{E}}^{2}{{\left\| {\mathbf{\bar{b}}} \right\|}^{2}} \right]$, ${{\psi }_{2}}=\frac{2{{\beta }^{2}}}{KD}$. It can be observed that the exponential functions in the left-hand-side (LHS) of (\ref{eq:16b}) are convex, whereas the convexity of the right-hand-side (RHS) results in that the constraint is non-convex. To this end, we use the first-order Taylor expansion to provide a lower bound of the RHS, i.e., 
	\begin{align}
		\frac{{{\rho }_{w}}}{{{\lambda }_{w}}}-{{c}_{w}}\ge -\frac{{{\rho }_{w}}}{\bar{\lambda }_{w}^{2}}{{\lambda }_{w}}+\frac{2{{\rho }_{w}}}{{{{\bar{\lambda }}}_{w}}}-{{c}_{w}}. \label{eq:17}
	\end{align}
	By substituting (\ref{eq:17}) into (\ref{eq:16b}), \textbf{SP1} becomes a standard convex optimization issue that can be addressed via interior-point method (IPM). Thereafter, we round $\left\{ {{B}_{t}} \right\}_{t=1}^{T}$ into integer values for satisfying their physical meanings. 
	
	\subsection{Transmission Coefficient Optimization}
	
	To handle the inter-agent coupling induced by feature correlations, we propose to alternatively update the transmission coefficient of each agent $k$, then the subproblem for optimizing ${{\nu }_{k}}$ is expressed as
	\begin{subequations}
		\begin{equation}
			\begin{aligned}
				\textbf{SP2}(k): \underset{{{\nu }_{k}}}{\mathop{\min }}\,{{\varpi }_{1,k}}{{\left| {{\nu }_{k}} \right|}^{2}}+2\operatorname{Re}\left\{ {{\varpi }_{2,k}}{{\nu }_{k}} \right\}, \label{eq:18a}
			\end{aligned}
		\end{equation}
		\vspace{-5mm}
		\begin{align}
			\mbox{s.t.}\ 
			&{{\left| {{\nu }_{k}} \right|}^{2}}\le \frac{{{P}_{\text{A}}}}{{{\beta }^{2}}J}, \label{eq:18b}\\
			&\text{tr}\big\{ \mathbf{\bar{\Phi }}_{\text{a}}^{\text{H}}\big[ {{\beta }^{2}}J\big( \mathbf{h}_{\text{AR},k}^{*}\mathbf{h}_{\text{AR},k}^{\text{T}}{{\left| {{\nu }_{k}} \right|}^{2}}+2\operatorname{Re}\left\{ {{\mathbf{\Xi }}_{1,k}}{{\nu }_{k}} \right\}+{{\mathbf{\Xi }}_{2,k}} \big)\nonumber\\
			&\qquad\qquad\qquad\qquad\qquad+\sigma _{\text{R}}^{2}{{\mathbf{I}}_{N}} \big]{{{\mathbf{\bar{\Phi }}}}_{\text{a}}} \big\}\le {{P}_{\text{R}}}, \label{eq:18c}
		\end{align} 
	\end{subequations}
    where ${{\varpi }_{1,k}}={{\mathbf{\bar{b}}}^{\text{H}}}\mathbf{\bar{h}}_{k}^{\text{*}}\mathbf{\bar{h}}_{k}^{\text{T}}\mathbf{\bar{b}}$, ${{\varpi }_{2,k}}=\sum\nolimits_{{k}'\in \mathcal{K}\backslash \left\{ k \right\}}{{{u}_{k,{k}'}}{{{\mathbf{\bar{b}}}}^{\text{H}}}\mathbf{\bar{h}}_{{{k}'}}^{\text{*}}\mathbf{\bar{h}}_{k}^{\text{T}}\mathbf{\bar{b}}\nu _{{{k}'}}^{*}}-\sum\nolimits_{{k}'\in \mathcal{K}}{{{u}_{k,{k}'}}\mathbf{\bar{h}}_{k}^{\text{T}}\mathbf{\bar{b}}}$, ${{\mathbf{\Xi }}_{1,k}}=\sum\nolimits_{{k}'\in \mathcal{K}\backslash \left\{ k \right\}}{{{u}_{k,{k}'}}\mathbf{h}_{\text{AR},k}^{\text{*}}\mathbf{h}_{\text{AR},{k}'}^{\text{T}}\bar{\nu }_{{{k}'}}^{*}}$, and ${{\mathbf{\Xi }}_{2,k}}=\sum\nolimits_{p\in \mathcal{K}\backslash \left\{ k \right\}}{\sum\nolimits_{{p}'\in \mathcal{K}\backslash \left\{ k \right\}}{{{u}_{p,{p}'}}\bar{\nu }_{p}^{*}{{{\bar{\nu }}}_{{{p}'}}}\mathbf{h}_{\text{AR},p}^{*}\mathbf{h}_{\text{AR},{p}'}^{\text{T}}}}$ are intermediate variables that do not relate to ${{\nu }_{k}}$. We can observe that the objective in (\ref{eq:18a}) and the LHSs of constraints (\ref{eq:18b})-(\ref{eq:18c}) are all quadratic functions w.r.t. ${{\nu }_{k}}$, hence \textbf{SP2}$\left( k \right)$ is convex and can be directly solved via IPM. 
    
    To gain more insights, we consider a special case in which the feature correlations among agents do not exist, namely ${{u}_{k,{k}'}}=0,\forall k\ne {k}'$ and ${{u}_{k,k}}=1,\forall k$, then the transmission coefficient optimization subproblem reduces to 
    \begin{subequations}
    	\begin{equation}
    		\begin{aligned}
    			\textbf{SP2.1}: \underset{\left\{ {{\nu }_{k}} \right\}_{k=1}^{K}}{\mathop{\min }}\,\sum\limits_{k\in \mathcal{K}}{{{\left| {{\nu }_{k}}\mathbf{\bar{h}}_{k}^{\text{T}}\mathbf{\bar{b}}-1 \right|}^{2}}}, \label{eq:19a}
    		\end{aligned}
    	\end{equation}
    	\vspace{-5mm}
    	\begin{align}
    		\mbox{s.t.}\ 
    		&{{\left| {{\nu }_{k}} \right|}^{2}}\le \frac{{{P}_{\text{A}}}}{{{\beta }^{2}}J},\forall k, \label{eq:19b}\\
    		&\sum\nolimits_{k\in \mathcal{K}}{{{\left| {{\nu }_{k}} \right|}^{2}}\text{tr}\left( \mathbf{\bar{\Phi }}_{\text{a}}^{\text{H}}\mathbf{h}_{\text{AR},k}^{*}\mathbf{h}_{\text{AR},k}^{\text{T}}{{{\mathbf{\bar{\Phi }}}}_{\text{a}}} \right)}\le {{\tilde{P}}_{\text{R}}}, \label{eq:19c}
    	\end{align} 
    \end{subequations}
	where ${{\tilde{P}}_{\text{R}}}=\frac{{{P}_{\text{R}}}-\text{tr}\left( \sigma _{\text{R}}^{2}\mathbf{\bar{\Phi }}_{\text{a}}^{\text{H}}{{{\mathbf{\bar{\Phi }}}}_{\text{a}}} \right)}{{{\beta }^{2}}J}$. In this case, we can derive a closed-form solution for the optimal $\left\{ {{\nu }_{k}} \right\}_{k=1}^{K}$ as below
	\begin{align}
		{{\nu }_{k}}=&\min \left\{ \frac{\left| \mathbf{\bar{h}}_{k}^{\text{T}}\mathbf{\bar{b}} \right|}{{{\left| \mathbf{\bar{h}}_{k}^{\text{T}}\mathbf{\bar{b}} \right|}^{2}}+\kappa \text{tr}\left( \mathbf{\bar{\Phi }}_{\text{a}}^{\text{H}}\mathbf{h}_{\text{AR},k}^{*}\mathbf{h}_{\text{AR},k}^{\text{T}}{{{\mathbf{\bar{\Phi }}}}_{\text{a}}} \right)},\sqrt{\frac{{{P}_{\text{A}}}}{{{\beta }^{2}}J}} \right\}\nonumber\\
		&\qquad\qquad\qquad\qquad\qquad\qquad\times {{\text{e}}^{-\text{j}\angle \mathbf{\bar{h}}_{k}^{\text{T}}\mathbf{\bar{b}}}},\ \forall k, \label{eq:20}
	\end{align}
    where $\kappa$ is the non-negative Lagrange multiplier associated with (\ref{eq:19c}). The derivation is directly from the KKT conditions of \textbf{SP2.1}, which is omitted for simplification. (\ref{eq:20}) reveals that the optimal agent transmission coefficient follows an effective-channel-inversion structure, i.e., if the system power budgets are adequately large, then ${{\nu }_{k}}=\frac{1}{\mathbf{\bar{h}}_{k}^{\text{T}}\mathbf{\bar{b}}}$ is for offsetting the misalignment error. Different from conventional AirComp \cite{22}, a regularization term is imposed by the amplification power constraint at the hybrid RIS. 
	
	\subsection{Receiving Beamforming Optimization}
	
	By fixing $\left\{ {{B}_{t}} \right\}_{t=1}^{T},\left\{ {{\nu }_{k}} \right\}_{k=1}^{K}$ and $\mathbf{\Phi }$, the subproblem for optimizing $\mathbf{b}$ is recast as
	\begin{align}
		\textbf{SP3}: \underset{\mathbf{b}}{\mathop{\min }}\,{{\mathbf{b}}^{\text{H}}}\mathbf{\Pi b}-2\operatorname{Re}\left\{ {{\bm{\omega} }^{\text{H}}}\mathbf{b} \right\}, \label{eq:21}
	\end{align}
	where $\mathbf{\Pi }=\sum\nolimits_{k\in \mathcal{K}}{\sum\nolimits_{{k}'\in \mathcal{K}}{{{\beta }^{2}}J{{u}_{k,{k}'}}\bar{\nu }_{k}^{*}{{{\bar{\nu }}}_{{{k}'}}}\mathbf{\bar{h}}_{k}^{\text{*}}\mathbf{\bar{h}}_{{{k}'}}^{\text{T}}}}+\sigma _{\text{R}}^{2}\mathbf{H}_{\text{RE}}^{\text{*}}\mathbf{\bar{\Phi }}_{\text{a}}^{\text{H}}{{\mathbf{\bar{\Phi }}}_{\text{a}}}\mathbf{H}_{\text{RE}}^{\text{T}}\!+\!\sigma _{\text{E}}^{2}{{\mathbf{I}}_{M}}$, $\bm{\omega} =\sum\nolimits_{k\in \mathcal{K}}{\sum\nolimits_{{k}'\in \mathcal{K}}{{{\beta }^{2}}J{{u}_{k,{k}'}}\bar{\nu }_{k}^{*}\mathbf{\bar{h}}_{k}^{\text{*}}}}$. Evidently, \textbf{SP3} is an unconstrained convex optimization problem, and we can derive the optimal receiving beamforming by setting the first derivative of the objective function to zero. Then, we have
    \begin{align}
    	\mathbf{\Pi b}-\bm{\omega} =0\Rightarrow \mathbf{b}={{\mathbf{\Pi }}^{-1}}\bm{\omega}. \label{eq:22}
    \end{align}

    After performing beamforming to obtain received signal ${{\mathbf{y}}_{t}}$, the EN should execute a signal processing algorithm to detect ${{\mathbf{x}}_{t}}$, thereby completing global feature reconstruction via (\ref{eq:6}). For equation ${{\mathbf{y}}_{t}}={{\mathbf{P}}_{t}}{{\mathbf{x}}_{t}}$, the column number of ${{\mathbf{P}}_{t}}$, ${{I}_{t}}$, may exceed the row number $J$, as vector quantization generally demands a large number of bits to ensure high precision \cite{19}. Fortunately, such an underdetermined equation can be solved by exploiting the sparsity of ${{\mathbf{x}}_{t}}$. Specifically, the number of non-zero elements in ${{\mathbf{x}}_{t}}$ is no greater than agent number $K$, and thus much smaller than the dimensionality ${{I}_{t}}$, rendering ${{\mathbf{x}}_{t}}$ highly sparse. Consequently, to recover ${{\mathbf{x}}_{t}}$ from ${{\mathbf{y}}_{t}}$, the EN can apply any compressive sensing algorithm, e.g., stagewise weak orthogonal matching pursuit (SWOMP) used in our experiments \cite{32}. 
    
    \subsection{RIS Reflection Beamforming Optimization}
    
    In this subsection, we seek to optimize the reflection matrix $\mathbf{\Phi }$ of the hybrid RIS, then the subproblem is rewritten as 
	\begin{subequations}
		\begin{equation}
			\begin{aligned}
				\textbf{SP4}: \underset{\mathbf{\Phi }}{\mathop{\min }}\,\text{tr}\left( \mathbf{\Phi }{{\mathbf{\Omega }}_{1}}{{\mathbf{\Phi }}^{\text{H}}}{{\mathbf{\Omega }}_{2}} \right)-2\operatorname{Re}\left\{ \text{tr}\left[ \mathbf{\Phi }{{\mathbf{\Omega }}_{3}} \right] \right\}, \label{eq:23a}
			\end{aligned}
		\end{equation}
		\vspace{-8mm}
		\begin{align}
			\mbox{s.t.}\ 
			\text{(\ref{eq:14d})}\sim \text{(\ref{eq:14f})},\nonumber
		\end{align} 
	\end{subequations}
    where ${{\mathbf{\Omega }}_{1}}=\mathbf{H}_{\text{RE}}^{\text{T}}\mathbf{\bar{b}}{{\mathbf{\bar{b}}}^{\text{H}}}\mathbf{H}_{\text{RE}}^{*}+\sigma _{\text{R}}^{2}{{\mathbf{I}}_{N}}$, ${{\mathbf{\Omega }}_{2}}=\sum\nolimits_{k\in \mathcal{K}}{\sum\nolimits_{{k}'\in \mathcal{K}}{{{\beta }^{2}}J{{u}_{k,{k}'}}\bar{\nu }_{k}^{*}{{{\bar{\nu }}}_{{{k}'}}}\mathbf{h}_{\text{AR},k}^{*}\mathbf{h}_{\text{AR},{k}'}^{\text{T}}}}$, and ${{\mathbf{\Omega }}_{3}}=\mathbf{H}_{\text{RE}}^{\text{T}}\mathbf{\bar{b}}\sum\nolimits_{k\in \mathcal{K}}{\sum\nolimits_{{k}'\in \mathcal{K}}{{{\beta }^{2}}J{{u}_{k,{k}'}}{{{\bar{\nu }}}_{{{k}'}}}\mathbf{h}_{\text{AR},{k}'}^{\text{T}}}}-\mathbf{H}_{\text{RE}}^{\text{T}}\mathbf{\bar{b}}{{\mathbf{\bar{b}}}^{\text{H}}}\sum\nolimits_{k\in \mathcal{K}}{\sum\nolimits_{{k}'\in \mathcal{K}}{{{\beta }^{2}}J{{u}_{k,{k}'}}\bar{\nu }_{k}^{*}{{{\bar{\nu }}}_{{{k}'}}}\mathbf{h}_{\text{AE},k}^{*}\mathbf{h}_{\text{AR},{k}'}^{\text{T}}}}$, while equality $\mathbf{\Phi }_{\text{a}}^{\text{H}}{{\mathbf{\Phi }}_{\text{a}}}={{\mathbf{\Phi }}^{\text{H}}}\mathbf{\Phi }-\mathbf{\Phi }_{\text{p}}^{\text{H}}{{\mathbf{\Phi }}_{\text{p}}}$ is employed to replace ${{\mathbf{\Phi }}_{\text{a}}}$ with $\mathbf{\Phi }$ in the objective function, since $\mathbf{\Phi }_{\text{p}}^{\text{H}}{{\mathbf{\Phi }}_{\text{p}}}$ is a constant. Afterwards, we define three $N$-dimensional vectors $\mathbf{d}$, ${{\mathbf{d}}_{\text{a}}}$, and ${{\mathbf{d}}_{\text{p}}}$, whose elements are the diagonal entries of $\mathbf{\Phi }$, ${{\mathbf{\Phi }}_{\text{a}}}$, and ${{\mathbf{\Phi }}_{\text{p}}}$, respectively. The expression of \textbf{SP4} is simplified as 
    \begin{subequations}
    	\begin{equation}
    		\begin{aligned}
    			\textbf{SP4.1}: \underset{\mathbf{d}}{\mathop{\min }}\,{{\mathbf{d}}^{\text{H}}}{{\mathbf{R}}_{1}}\mathbf{d}-2\operatorname{Re}\left\{ \mathbf{r}_{1}^{\text{H}}\mathbf{d} \right\}, \label{eq:24a}
    		\end{aligned}
    	\end{equation}
    	\vspace{-8mm}
    	\begin{align}
    		\mbox{s.t.}\ 
    		&\left| {{\left[ {{\mathbf{d}}_{\text{p}}} \right]}_{n}} \right|=1,\text{ }\forall n\in {{\mathcal{N}}_{\text{p}}}, \label{eq:24b}\\
    		&\text{tr}\left[ \mathbf{d}_{\text{a}}^{\text{H}}{{\mathbf{R}}_{2}}{{\mathbf{d}}_{\text{a}}} \right]\le {{P}_{\text{R}}}, \label{eq:24c}\\
    		&\text{(\ref{eq:14d})},\nonumber
    	\end{align} 
    \end{subequations}
    where ${{\mathbf{R}}_{1}}=\mathbf{\Omega }_{1}^{\text{T}}\odot {{\mathbf{\Omega }}_{2}}$, the elements in ${{\mathbf{r}}_{1}}$ are the diagonal entries of $\mathbf{\Omega }_{3}^{\text{H}}$, and ${{\mathbf{R}}_{2}}={{\beta }^{2}}J\sum\nolimits_{k\in \mathcal{K}}{\sum\nolimits_{{k}'\in \mathcal{K}}{{{u}_{k,{k}'}}\bar{\nu }_{k}^{*}{{{\bar{\nu }}}_{{{k}'}}}\left( \mathbf{h}_{\text{AR},k}^{*}\mathbf{h}_{\text{AR},{k}'}^{\text{T}} \right)\odot {{\mathbf{I}}_{N}}}}+\sigma _{\text{R}}^{2}{{\mathbf{I}}_{N}}$. Although \textbf{SP4.1} can be handled by semidefinite relaxation (SDR) by converting the variable into $\mathbf{\tilde{D}}=\mathbf{\tilde{d}}{{\mathbf{\tilde{d}}}^{\text{H}}}$ with $\mathbf{\tilde{d}}={{[ {{\mathbf{d}}^{\text{T}}},1 ]}^{\text{T}}}$, the computational complexity is prohibit, i.e., $\mathcal{O}( {{( N+1 )}^{7}} )$, for directly optimizing matrix $\mathbf{\tilde{D}}$ \cite{21,33}. Towards this end, we propose a low-complexity iterative approach in the sequel, enabling efficient update of all variables in closed-form. 
    
    We observe that $\mathbf{d}$ is uniquely determined by two factors, i.e., amplitude matrix of active elements ${{\mathbf{\Lambda }}_{\text{a}}}=\text{diag}\left( \mathbf{1}_{N\times 1}^{\text{a}}\odot {{\left[ \left| {{\phi }_{1}} \right|,\ldots ,\left| {{\phi }_{N}} \right| \right]}^{\text{T}}} \right)$ and phase shift vector of the hybrid RIS $\bm{\alpha} ={{\left[ {{\text{e}}^{j{{\alpha }_{1}}}},\ldots ,{{\text{e}}^{j{{\alpha }_{N}}}} \right]}^{\text{T}}}$, together with 
    \begin{align}
    	\mathbf{d}={{\mathbf{\Lambda }}_{\text{a}}}\bm{\alpha} +\text{diag}\left( \mathbf{1}_{N\times 1}^{\text{p}} \right)\bm{\alpha}. \label{eq:25}
    \end{align}
    Motivated by this, we derive an upper bound for the objective in (\ref{eq:24a}) w.r.t. ${{\mathbf{\Lambda }}_{\text{a}}}$ and $\bm{\alpha} $ as follows.
    \begin{align}
    	&{{\mathbf{d}}^{\text{H}}}{{\mathbf{R}}_{1}}\mathbf{d}-2\operatorname{Re}\left\{ {{\mathbf{r}}^{\text{H}}}\mathbf{d} \right\}\overset{\left( a \right)}{\mathop{\le }}\,{{\sigma }_{\max }}\left( {{\mathbf{R}}_{1}} \right)\mathbf{d}_{\text{a}}^{\text{H}}{{\mathbf{d}}_{\text{a}}}-2\operatorname{Re}\left\{ \mathbf{\bar{r}}_{2}^{\text{H}}\mathbf{d} \right\}\!+\!c\nonumber\\
    	&\!=\!{{\sigma }_{\max }}\!\left( {{\mathbf{R}}_{1}} \right)\text{tr}\!\left( {{\mathbf{\Lambda }}_{\text{a}}}{{\mathbf{\Lambda }}_{\text{a}}} \right)\!-\!2\operatorname{Re}\left\{ \mathbf{\bar{r}}_{2}^{\text{H}}\left( {{\mathbf{\Lambda }}_{\text{a}}}\!+\!\text{diag}\left( \mathbf{1}_{N\times 1}^{\text{p}} \right) \right)\bm{\alpha}  \right\}\!+\! c, \label{eq:26}
    \end{align}
    where ${{\sigma }_{\max }}\left( {{\mathbf{R}}_{1}} \right)$ indicates the maximum eigenvalue of ${{\mathbf{R}}_{1}}$, ${{\mathbf{\bar{r}}}_{2}}={{\mathbf{r}}_{1}}+\left( {{\sigma }_{\max }}\left( {{\mathbf{R}}_{1}} \right){{\mathbf{I}}_{N}}-\mathbf{R}_{1}^{\text{H}} \right)\mathbf{\bar{d}}$, $c={{\sigma }_{\max }}\left( {{\mathbf{R}}_{1}} \right){{N}_{\text{p}}}+{{\mathbf{\bar{d}}}^{\text{H}}}\left( {{\sigma }_{\max }}\left( {{\mathbf{R}}_{1}} \right){{\mathbf{I}}_{N}}-{{\mathbf{R}}_{1}} \right)\mathbf{\bar{d}}$. $\left( a \right)$ follows the inequality in \cite[Eq.~(35)]{34} as well as ${{\mathbf{d}}^{\text{H}}}\mathbf{d}=\mathbf{d}_{\text{a}}^{\text{H}}{{\mathbf{d}}_{\text{a}}}+{{N}_{\text{p}}}$. It is readily inferred from (\ref{eq:26}) that the optimal $\bm{\alpha}$ is updated as
	\begin{align}
		{{\alpha }_{n}}={{{\left[ {{{\mathbf{\bar{r}}}}_{2}} \right]}_{n}}}/{\left| {{\left[ {{{\mathbf{\bar{r}}}}_{2}} \right]}_{n}} \right|},\ \forall n.  \label{eq:27}
	\end{align}
    By substituting (\ref{eq:27}) into (\ref{eq:26}), whilst defining ${{\mathbf{q}}_{{{\mathbf{\Lambda }}_{\text{a}}}}}\in {{\mathbb{R}}^{{{N}_{\text{a}}}\times 1}}$ as the vector acquired by stacking the ${{N}_{\text{a}}}$ non-zero diagonal entries of ${{\mathbf{\Lambda }}_{\text{a}}}$, we establish the following problem for determining ${{\mathbf{q}}_{{{\mathbf{\Lambda }}_{\text{a}}}}}$.
    \begin{subequations}
    	\begin{equation}
    		\begin{aligned}
    			\textbf{SP4.2}: \underset{{{\mathbf{q}}_{{{\mathbf{\Lambda }}_{\text{a}}}}}}{\mathop{\min }}\,{{\sigma }_{\max }}\left( {{\mathbf{R}}_{1}} \right)\mathbf{q}_{{{\mathbf{\Lambda }}_{\text{a}}}}^{\text{T}}{{\mathbf{q}}_{{{\mathbf{\Lambda }}_{\text{a}}}}}-2\mathbf{\bar{r}}_{3}^{\text{T}}{{\mathbf{q}}_{{{\mathbf{\Lambda }}_{\text{a}}}}}, \label{eq:28a}
    		\end{aligned}
    	\end{equation}
    	\vspace{-8mm}
    	\begin{align}
    		\mbox{s.t.}\ 
    		&\mathbf{q}_{{{\mathbf{\Lambda }}_{\text{a}}}}^{\text{T}}{{\mathbf{R}}_{3}}{{\mathbf{q}}_{{{\mathbf{\Lambda }}_{\text{a}}}}}\le {{P}_{\text{R}}}, \label{eq:28b}
    	\end{align} 
    \end{subequations}
    where ${{\mathbf{\bar{r}}}_{3}}={{\left[ \left| {{\left[ {{{\mathbf{\bar{r}}}}_{2}} \right]}_{n}} \right|:n\in {{\mathcal{N}}_{\text{a}}} \right]}^{\text{T}}}$, ${{\mathbf{R}}_{3}}=\text{diag}\left( {{\left[ {{\mathbf{R}}_{2}} \right]}_{n,n}}:n\in {{\mathcal{N}}_{\text{a}}} \right)$. The optimal solution for \textbf{SP4.2} can be obtained through the KKT condition, given by
    \begin{align}
    	{{\mathbf{q}}_{{{\mathbf{\Lambda }}_{\text{a}}}}}={{\left( {{\sigma }_{\max }}\left( {{\mathbf{R}}_{1}} \right){{\mathbf{I}}_{{{N}_{\text{a}}}}}+\varsigma {{\mathbf{R}}_{3}} \right)}^{-1}}{{\mathbf{\bar{r}}}_{3}}, \label{eq:29}
    \end{align}
    where $\varsigma$ is the multiplier associated with (\ref{eq:28b}), whose value is calculated by 
    \begin{align}
    	\varsigma=
    	\left\{\begin{array}{cl}
    		&\!\!\!\!\!\!\!\!\!0,\text{ }\sum\nolimits_{n=1}^{{{N}_{\text{a}}}}{{{\left( {{\left[ {{{\mathbf{\bar{r}}}}_{3}} \right]}_{n}} \right)}^{2}}{{\left[ {{\mathbf{R}}_{3}} \right]}_{n,n}}}\le {{P}_{\text{R}}}\sigma _{\max }^{2}\left( {{\mathbf{R}}_{1}} \right),\\
    		&\!\!\!\!\!\!\!\!\!\text{Solution of }f\left( \varsigma  \right)={{P}_{\text{R}}},\text{ otherwise}, \\
    	\end{array}\right.  \label{eq:30}
    \end{align}
    where $f\left( \varsigma  \right)=\sum\limits_{n=1}^{{{N}_{\text{a}}}}{\frac{{{\left( {{\left[ {{{\mathbf{\bar{r}}}}_{3}} \right]}_{n}} \right)}^{2}}{{\left[ {{\mathbf{R}}_{3}} \right]}_{n,n}}}{{{\left( {{\sigma }_{\max }}\left( {{\mathbf{R}}_{1}} \right)+\varsigma {{\left[ {{\mathbf{R}}_{3}} \right]}_{n,n}} \right)}^{2}}}}$. 
    
    To summarize, the RIS reflection beamforming is optimized by alternatively updating $\bm{\alpha}$, ${{\mathbf{q}}_{{{\mathbf{\Lambda }}_{\text{a}}}}}$, and $\mathbf{d}$ via (\ref{eq:27}), (\ref{eq:29}), and (\ref{eq:25}), respectively, and the procedure continues until the difference of the objective value (\ref{eq:24a}) in two adjacent iterations is sufficiently small.
    
    \subsection{Joint Optimization Algorithm}
    
    \begin{algorithm}[t]
    	\caption{Joint feature Quantization and Active-Passive Beamforming design (JQAPB) algorithm } \label{alg:1}
    	\begin{algorithmic}[1]
    		\STATE \textbf{Input:} Channel information $\left\{ {{\mathbf{h}}_{\text{AE},k}},{{\mathbf{h}}_{\text{AR},k}} \right\}_{k=1}^{K},{{\mathbf{H}}_{\text{RE}}}$, hybrid RIS configuration ${{\mathcal{N}}_{\text{a}}},{{\mathcal{N}}_{\text{p}}}$, features’ importance indicators and variances $\left\{ {{\rho }_{w}},{{c}_{w}} \right\}_{w=1}^{W}$, number of available bits $B$, power budgets ${{P}_{\text{A}}},{{P}_{\text{R}}}$. 
    		\STATE \textbf{Initialize:} $\left\{ {{{\bar{\lambda }}}_{w}} \right\}_{w=1}^{W},\left\{ {{{\bar{\nu }}}_{k}} \right\}_{k=1}^{K},\mathbf{\bar{b}},\mathbf{\bar{\Phi }}$, convergence threshold ${{\Delta }_{\text{out}}},{{\Delta }_{\text{in}}}\ll 1$. 
    		\REPEAT
    		\STATE\textbf{Outer iteration:}
    		\STATE Optimize $\left\{ {{B}_{t}} \right\}_{t=1}^{T}$ by solving \textbf{SP1}.
    		\FOR{$k \in \mathcal{K}$}
    		\STATE Optimize ${{\nu }_{k}}$ by solving \textbf{SP2}$\left( k \right)$.
    		\ENDFOR
    		\STATE Optimize $\mathbf{b}$ using (\ref{eq:22}). 
    		\REPEAT
    		\STATE\textbf{Inner iteration:}
    		\STATE Update $\bm{\alpha}$, ${{\mathbf{q}}_{{{\mathbf{\Lambda }}_{\text{a}}}}}$, and $\mathbf{d}$ using (\ref{eq:27}), (\ref{eq:29}), and (\ref{eq:25}), respectively.
    		\UNTIL{Change of the objective value in (\ref{eq:24a}) is less than ${{\Delta }_{\text{in}}}$}
    		\STATE Optimize $\mathbf{\Phi }=\text{diag}\left( \mathbf{d} \right)$.
    		\STATE Calculate $\left\{ c_{w}^{\mathbf{e}} \right\}_{w=1}^{W}$ and $G=\sum\limits_{w=1}^{W}{\frac{{{\rho }_{w}}}{{{c}_{w}}+c_{w}^{\mathbf{e}}}}$.
    		\UNTIL{Change of the objective value in (\ref{eq:14a}) is less than ${{\Delta }_{\text{out}}}$} 
    		\STATE \textbf{Output:} $\left\{ {{B}_{t}} \right\}_{t=1}^{T},\left\{ {{\nu }_{k}} \right\}_{k=1}^{K},\mathbf{b},\mathbf{\Phi }$. 
    	\end{algorithmic}
    \end{algorithm}

	The developed Joint feature Quantization and Active-Passive Beamforming design (JQAPB) algorithm for addressing \textbf{P1} is outlined in \textbf{Algorithm 1}. We commence by initializing the optimization variables and setting convergence thresholds. The algorithm then iteratively updates the quantization bit allocation $\left\{ {{B}_{t}} \right\}_{t=1}^{T}$, transmission coefficients $\left\{ {{\nu }_{k}} \right\}_{k=1}^{K}$, receiving beamforming $\mathbf{b}$, and RIS reflection matrix $\mathbf{\Phi }$ following the procedures detailed in the preceding subsections. Finally, \textbf{Algorithm 1} converges to the sub-optimal solution of \textbf{P1} when the change of the objective value in (\ref{eq:14a}) is less than the predefined threshold. The optimized variables are leveraged to construct the codebooks for feature quantization and modulation, as well as coordinate the HRD-AirComp-based feature aggregation process, thereby boosting the edge inference performance.
    
    \textit{Theorem 3:} \textbf{Algorithm 1} monotonically converges to a stationary solution to \textbf{P1} within a finite number of iterations.
    
    \textit{Proof:} Please refer to Appendix C. $\qquad\qquad\qquad\qquad\ \ \blacksquare$
    
    The computational complexity of \textbf{Algorithm 1} is analyzed as follows. Suppose that the outer alternative optimization and inner RIS reflection updating require ${{I}_{\text{out}}}$ and ${{I}_{\text{in}}}$ iterations, respectively. Given that IPM possesses a complexity of $\mathcal{O}\left( {{\xi }^{3.5}} \right)$ with $\xi$ being the number of variables, the complexities for optimizing the four sets of variables are $\mathcal{O}\left( {{T}^{3.5}} \right)$, $\mathcal{O}\left( {{K}^{3.5}} \right)$, $\mathcal{O}\left( {{M}^{2}} \right)$, and $\mathcal{O}\left( {{I}_{\text{in}}}{{N}^{2}} \right)$, respectively. Accordingly, the total complexity of \textbf{Algorithm 1} is given by $\mathcal{O}\left( {{I}_{\text{out}}}\left( {{T}^{3.5}}+{{K}^{3.5}}+{{M}^{2}}+{{I}_{\text{in}}}{{N}^{2}} \right) \right)$. Compared with the conventional SDR-based RIS reflection solution, whose complexity is $\mathcal{O}\left( {{I}_{\text{out}}}\left( {{T}^{3.5}}+{{K}^{3.5}}+{{M}^{2}}+{{(N+1)}^{7}+I_{\text{Gauss}}} \right) \right)$ with $I^{\text{Guass}}$ being the number of candidate vectors generated via Gaussian randomization, the proposed JQAPB maintains a moderate complexity as $W$, $T$, and $N$ increase. It is worth noting that JQAPB is executed at the EN based on the inference requests and channel state information, with optimized transmission coefficients fed back to agents, thus does not impose additional local computational burden.
    
    \textit{Remark 4 (Extension to More General RIS Setups):} \textit{On the one hand}, the proposed scheme can be extended to scenarios involving multiple RISs by augmenting the channel $\mathbf{h}_k$ to account for both single-reflection and multi-reflection links induced by different RISs \cite{21}. In such cases, the reflection beamforming associated with different RISs can be optimized in an alternative manner, where each update follows the same procedure in Section V-E. \textit{On the other hand}, when discrete phase shifts and amplitudes are considered, they can be obtained by quantizing the optimized continuous reflection coefficients in $\mathbf{\Phi}$ to their nearest discrete levels. With sufficient control resolution and number of RIS elements $N$, the performance loss is typically negligible.
    
	\section{Performance Evaluation} \label{sec:evaluation}
	
	\subsection{Simulation Settings}
	
	In this section, experimental results are presented to evaluate the performance of the proposed HRD-AirComp scheme and JAQPB algorithm. We consider a 3D coordinate system, where the EN and hybrid RIS are located at (5, 0, 15) m and (0, 10, 15) m, respectively. All agents are randomly distributed in a circle area with center (25, 50, 0) m and radius of 20 m. We adopt the Rician channel model \cite{20,25,33} given by $\mathbf{h}=\sqrt{{{10}^{-3}}\cdot {{d}^{-\vartheta }}}\left( \sqrt{\frac{\chi }{1+\chi }}{{\mathbf{h}}^{\text{LoS}}}+\sqrt{\frac{1}{1+\chi }}{{\mathbf{h}}^{\text{NLoS}}} \right)$, where $d$ represents the distance, $\vartheta$ is the pathloss exponents, $\chi$ indicates the Rician factor, the line-of-sight (LoS) component ${{\mathbf{h}}^{\text{LoS}}}$ is determined by the steering vectors, and the non-LoS component ${{\mathbf{h}}^{\text{NLoS}}}$ is modeled as ${{\left[ {{\mathbf{h}}^{\text{NLoS}}} \right]}_{m}}\sim \mathcal{C}\mathcal{N}\left( 0,1 \right)$. In our simulation, agent-EN, agent-RIS, and RIS-EN links possess different parameters $\left\{ {{\vartheta }_{\text{AE}}},{{\vartheta }_{\text{AR}}},{{\vartheta }_{\text{RE}}} \right\}$ and $\left\{ {{\chi }_{\text{AE}}},{{\chi }_{\text{AR}}},{{\chi }_{\text{RE}}} \right\}$, as listed in Table \ref{tab:I}. 
	
	\begin{table}[t]\footnotesize
		\centering
		\setlength{\abovecaptionskip}{0pt}    
		\setlength{\belowcaptionskip}{10pt}
		\caption{Simulation Parameters} \label{tab:I}
		\begin{threeparttable}
			\begin{tabular}{cc}
				\hline
				\textbf{Parameter}&\textbf{Value}\\
				\hline
				Number of antennas at the EN $M$ & 16 \cite{12}\\
				Total number of elements at the hybrid RIS $N$&64 \cite{33}\\
				Number of active elements at the hybrid RIS ${{N}_{\text{a}}}$&8 \cite{25}\\
				Pathloss exponents ${{\vartheta }_{\text{AE}}},{{\vartheta }_{\text{AR}}},{{\vartheta }_{\text{RE}}}$ & 3.7, 2.2, 2 \cite{25}\\
				Rician factors ${{\chi }_{\text{AE}}},{{\chi }_{\text{AR}}},{{\chi }_{\text{RE}}}$ & 0, 1, $\infty$ \cite{25}\\
				RIS noise variance $\sigma _{\text{R}}^{2}$ & -70 dBm \cite{33}\\
				Channel noise variance $\sigma _{\text{E}}^{2}$ & -80 dBm \cite{33}\\
				Block length $D$ & 20 \cite{17}\\
				Sequence length $J$ & 70 \cite{17}\\
				Block norm bound $\beta$ & 18.9 \cite{35}\\
				Constant related to detection MSE $\eta$ & 1 \cite{17}\\
				Agent power budget ${{P}_{\text{A}}}$ & 20 dBm \cite{18}\\
				Hybrid RIS power budget ${{P}_{\text{R}}}$ & 23 dBm \cite{25}\\
				Convergence threshold ${{\Delta }_{\text{out}}},{{\Delta }_{\text{in}}}$ & 10$^\text{-5}$, 10$^\text{-6}$ \cite{25}\\
				\hline
			\end{tabular}
		\end{threeparttable}			
	\end{table} 
    
    To comprehensively evaluate the performance, we simulate two representative types of inference tasks:
    
    \textit{1) Linear Classification on Gaussian Mixture Dataset:} The global features $\mathbf{f}$ are sampled from a Gaussian mixture distribution $\frac{1}{L}\sum\nolimits_{l=1}^{L}{\mathcal{N}\left( {{\bm{\mu} }_{l}},\mathbf{C} \right)}$ with dimension $W=100$ and class number $L=20$. The local features are generated by ${{\mathbf{f}}_{k}}=\mathbf{f}+{{\mathbf{w}}_{k}}$, where ${{\mathbf{w}}_{k}}$ signifies the feature extraction noise of agent $k$, following a Gaussian distribution $\mathcal{N}\left( \mathbf{0},\sigma _{\text{F}}^{2}{{\mathbf{I}}_{W}} \right)$ with variance $\sigma _{\text{F}}^{2}=0.5$. After obtaining the reconstructed features $\mathbf{\hat{f}}$, the EN calculates the classification result as $\arg \underset{l}{\mathop{\min }}\,{( \mathbf{\hat{f}}-{{\bm{\mu} }_{l}} )^{\text{T}}}{( \mathbf{C}+\sigma _{\text{F}}^{2}{{\mathbf{I}}_{W}} )^{-1}}( \mathbf{\hat{f}}-{{\bm{\mu} }_{l}} )$. We set the number of agents $K=24$ and available quantization bits $B=40$ in default. 
    
    \textit{2) Multi-View Object Recognition on ModelNet Dataset:} In this kind of task, $K=12$ agents collect multi-view images of a common object, and the EN aims to classify the object from $L=40$ classes. This multi-view perception setting is closely related to practical applications such as autonomous driving and low-altitude economy, where objects (e.g., person, vehicle, and aircraft) are observed from distributed viewpoints and collaboratively recognized to support environment understanding and decision making. We train a lightweight MobileNet-v2 on the ModelNet dataset, and deploy the layers before the linear classifier at each agent for feature extraction \cite{35}. The resultant feature dimension $W$ is 1280, and the number of bits is set as $B=512$. The EN feeds the reconstructed $\mathbf{\hat{f}}$ into the linear classifier to obtain the classification result. The inference accuracies presented in Section VI-C are on the test data of ModelNet. 
    
    Moreover, we employ the Lloyd algorithm \cite{28} to design the Grassmannian codebook $\left\{ {{\mathbf{Q}}_{t}} \right\}$ based on the optimized bit allocation. The elements of the modulation codebook $\left\{ {{\mathbf{P}}_{t}} \right\}$ are drawn from complex Bernoulli distribution, i.e., ${{\left[ {{\mathbf{P}}_{t}} \right]}_{j,i}}\in \left\{ {\left( \pm 1\pm \text{j} \right)}/{\sqrt{2}} \right\}$ \cite{36}. The compressive sensing algorithm used for detecting the sparse signal ${{\mathbf{x}}_{t}}$ is SWOMP \cite{32}, primarily due to its advantage of not requiring prior knowledge on the sparsity level. 
    
    \begin{figure}[t] \centering
    	\subfigure[Convergence of the outer iterations under different $K$.] { 
    		{\includegraphics[width=4.1cm]{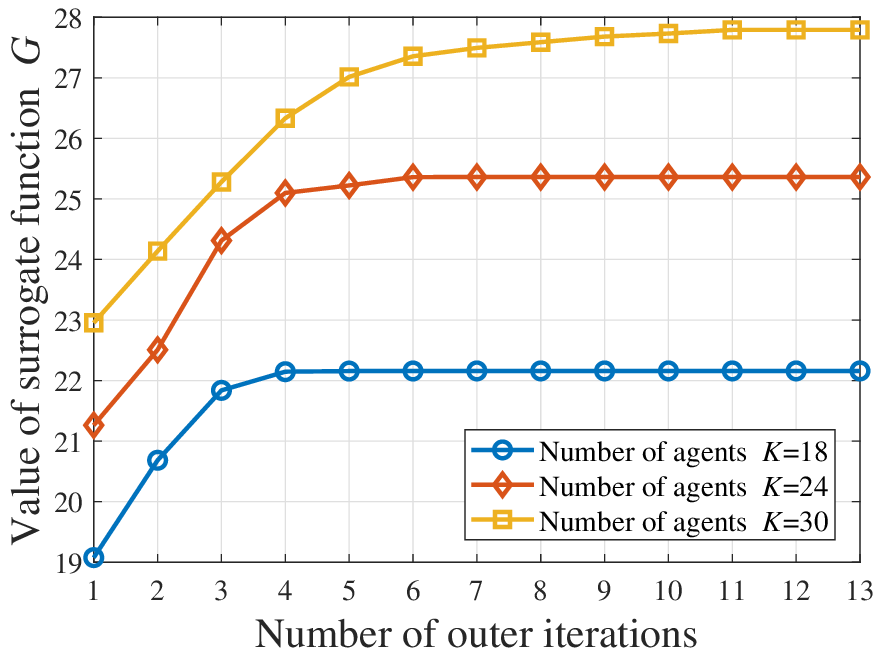}} 
    	}     
    	\subfigure[Convergence of inner iterations.] { 
    		{\includegraphics[width=4.1cm]{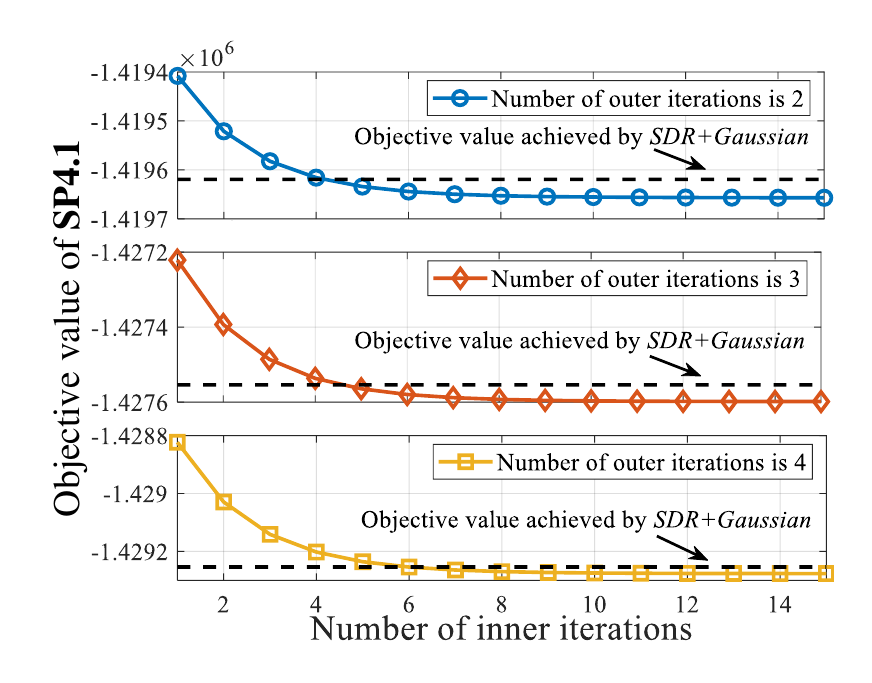}} 
    	}
    	\vspace{0mm}
    	\caption{Convergence of the proposed JQAPB algorithm.}     
    	\label{fig:3}     
    \end{figure}
    
    \begin{table}[t]\footnotesize
    	\centering
    	\caption{Implementation Time Comparison of Different RIS Reflection Optimization Methods}\label{tab:II}
    	\vspace{2.5 mm}
    	\begin{tabular}{cccc}
    		\Xhline{1\arrayrulewidth}
    		\multirow{2}{*}{Methods} & \multicolumn{3}{c}{Number of elements at the hybrid RIS}\\\cline{2-4}
    		& $N=$ 64 & $N=$ 128 & $N=$ 256  \\ 
    		\hline
    		JQAPB, proposed	&  \multirow{2}{*}{0.0246 s} & \multirow{2}{*}{0.0504 s} & \multirow{2}{*}{0.3176 s}\\
    		(solve \textbf{SP4.1} once)&&&\\
    		\textit{SDR+Gaussian}	&  \multirow{2}{*}{4.4594 s} & \multirow{2}{*}{5.8914 s} & \multirow{2}{*}{21.0890 s}\\
    		(solve \textbf{SP4.1} once)&&&\\
    		JQAPB, proposed	&  \multirow{2}{*}{0.6175 s} & \multirow{2}{*}{1.2390 s} & \multirow{2}{*}{7.2310 s}\\
    		(solve \textbf{P1} completely)&&&\\
    		\textit{SDR+Gaussian} &  \multirow{2}{*}{50.451 s} & \multirow{2}{*}{140.822 s} & \multirow{2}{*}{352.210 s} \\
    		(solve \textbf{P1} completely)&&&\\
    		\Xhline{1\arrayrulewidth}   	
    	\end{tabular}
    	%\vspace{-0.25in}
    \end{table}

    The following baselines are invoked for comparison: 
    
    \textit{1) Massive Digital (MD)-AirComp \cite{17}:} This state-of-the-art digital AirComp scheme employs a common codebook, optimized using the K-means++ algorithm, to quantize all feature blocks with uniform bit allocation. Moreover, due to the absence of theoretical analysis on the inference performance in \cite{17}, truncated channel inversion and minimum-MSE beamforming are adopted by agents and EN during over-the-air feature aggregation, respectively. Meanwhile, hybrid RIS deployment is not considered in this scheme. 
    
    \textit{2) One-Bit Digital Aggregation (OBDA) \cite{13}:} In this scheme, each dimension of the local feature vector is quantized using a single bit, where the sign is mapped to a BPSK symbol for transmission. The EN reconstructs the aggregated feature vector by extracting the sign of the received signal. This scheme adopts the same design for transmission coefficients and receiving beamforming as in \textit{MD-AirComp}. 
    
    \textit{3) Full Power Transmission:} Each agent leverages the full power budget to transmit signals by setting ${{\nu }_{k}}=\sqrt{\frac{{{P}_{\text{A}}}}{{{\beta }^{2}}J}}{{\text{e}}^{-\text{j}\angle \mathbf{\bar{h}}_{k}^{\text{T}}\mathbf{\bar{b}}}},\forall k$, whilst the other optimization variables, i.e., bit allocation, receiving beamforming, and RIS reflection, are optimized by our JQAPB algorithm. 
    
    \textit{4) SDR+Gaussian \cite{21,33}:} This baseline replaces the RIS reflection optimization in JQAPB with an SDR-based method. Concretely, \textbf{SP4.1} is solved by optimizing the $\left( N+1 \right)\times \left( N+1 \right)$ matrix $\mathbf{\tilde{D}}$ followed by recovering a rank-one solution via Gaussian randomization. To ensure satisfactory performance, 10$^\text{4}$ random vectors are generated during the randomization process. 
    
    \textit{5) Perfect Feature Aggregation (PFA):} This benchmark assumes that the transmission signals of all agents are perfectly aggregated at the EN, free from both misalignment error and channel noise. As such, the EN can directly use ${{\mathbf{x}}_{t}}=\sum\nolimits_{k\in \mathcal{K}}{{{\beta }_{k,t}}{{\mathbf{x}}_{k,t}}}$ to reconstruct the global features. 
    
    \textit{6) Ideal Edge Inference:} In this setting, the EN receives the average of local features, i.e., $\mathbf{f}=\frac{1}{K}\sum\nolimits_{k\in \mathcal{K}}{{{\mathbf{f}}_{k}}}$, without any quantization or transmission distortion. $\mathbf{f}$ is then fed into the EN-side AI model to yield the inference result. This benchmark serves as an upper bound on the inference performance.
    
    \subsection{Performance Evaluation for Linear Classification}
    
    \begin{figure*}[t]
    	\begin{tabular}{ccc}
    		\begin{minipage}[t]{0.31\linewidth}\centering
    			\includegraphics[width = 4.8cm]{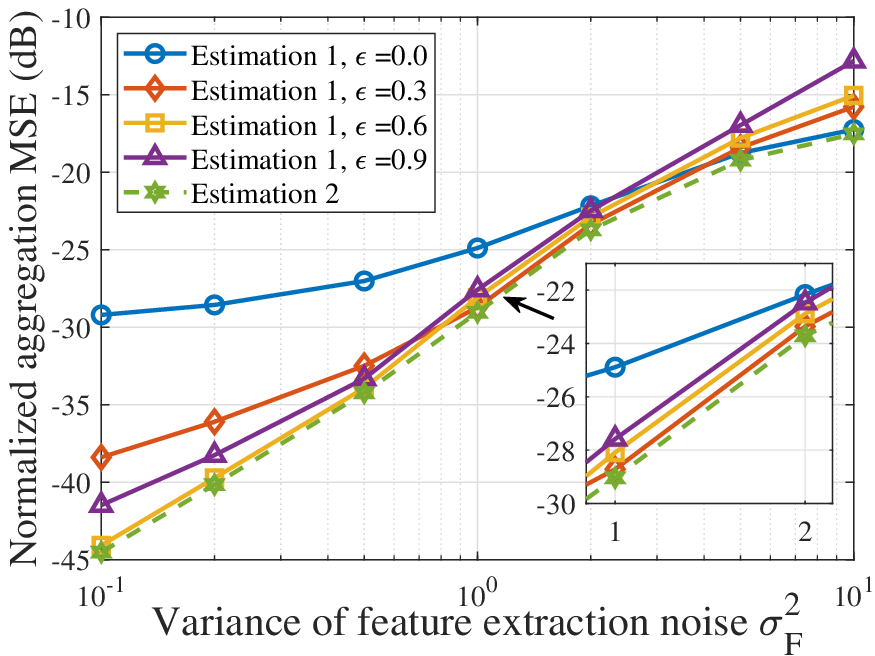}
    			\vspace{1mm}
    			\caption{Aggregation MSE versus feature extraction variance with different estimations of correlation matrix $\mathbf{U}$. }
    			\label{fig:4}   
    		\end{minipage}\hspace{0.3 cm}
    		\begin{minipage}[t]{0.31\linewidth}\centering
    			\includegraphics[width = 4.8cm]{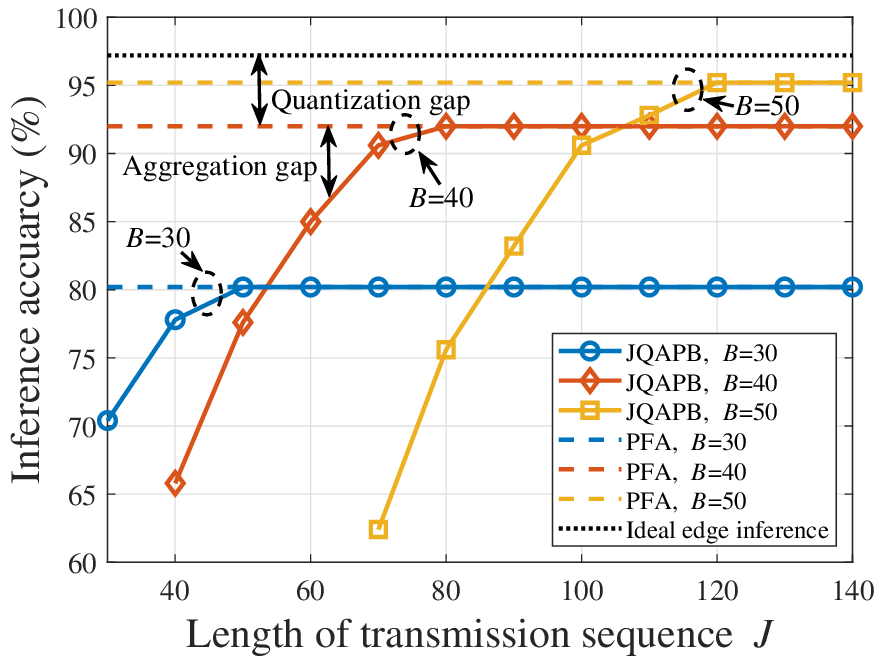}
    			\vspace{1mm}
    			\caption{Inference accuracy under different sequence length $J$ and available bit number $B$.}
    			\label{fig:5} 
    		\end{minipage}\hspace{0.3 cm}
    		\begin{minipage}[t]{0.31\linewidth}\centering
    			\includegraphics[width = 4.8cm]{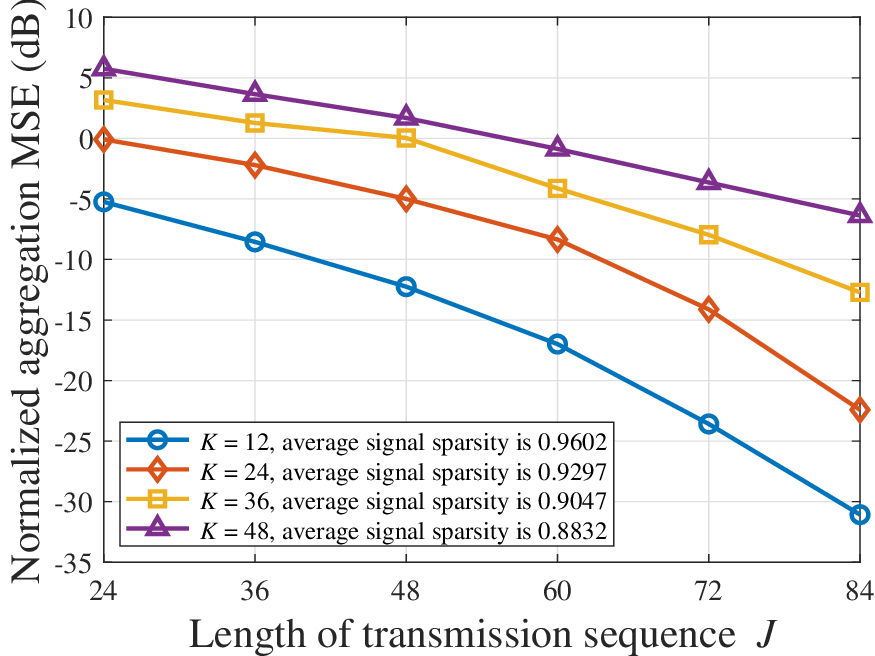}
    			\vspace{1mm}
    			\caption{Aggregation MSE versus under different signal sparsity and $J$.}
    			\label{fig:sparsity_impact} 
    		\end{minipage}
    		\vspace{0mm}	
    	\end{tabular}
    \end{figure*}

    Fig. \ref{fig:3} illustrates the convergence behavior of the proposed JQAPB algorithm. Specifically, Fig. \ref{fig:3} (a) presents the convergence of the outer iteration, where the four sets of variables are optimized in an alternating fashion. It can be observed that for different values of the agent number $K$, the surrogate function $G$ increases monotonically and converges within 10 outer iterations. Fig. \ref{fig:3} (b) further verifies the convergence of the inner iteration for optimizing the RIS reflection beamforming. We can see that the achieved objective value of \textbf{SP4.1} is consistently lower than that obtained by \textit{SDR+Gaussian}. This is because the proposed iterative approach leverages closed-form expressions to update the amplitude and phase shift of the hybrid RIS, thereby yielding higher-quality locally optimal solution compared to those recovered via Gaussian randomization. 

    In addition, Table \ref{tab:II} compares the implementation time of different RIS reflection beamforming methods for solving \textbf{SP4.1}. Compared to \textit{SDR+Gaussian}, the proposed iterative approach reduces the implementation time by approximately two orders of magnitude. This significant reduction in computational complexity is attributed to the fact that our method involves only arithmetic operations, avoiding the optimization of high-dimensional matrices and the generation of numerous random vectors required in \textit{SDR+Gaussian}. Furthermore, as the number of RIS elements $N$ increases, the runtime of our approach remains moderate, while that of \textit{SDR+Gaussian} becomes prohibitively high. 

    Subsequently, we investigate the necessity of incorporating transmission signal correlations (as elaborated in Assumption 3) into the AirComp beamforming design. We consider two estimations on the correlation matrix $\mathbf{U}$, where Estimation 1 sets $\mathbf{U}=\epsilon {{\mathbf{1}}_{K\times K}}+\left( 1-\epsilon  \right){{\mathbf{I}}_{K}}$ with $\epsilon $ being an empirical value on the degree of correlation. Notably, existing schemes often assume independent transmission signals, corresponding to the case of $\epsilon =0$ \cite{20,21,22,23,24}.  In Estimation 2, $\mathbf{U}$ is approximated based on the correlation among agents’ local features, i.e., ${{u}_{k,{k}'}}=\frac{1}{W}\sum\nolimits_{w=1}^{W}{{{[ {{{\mathbf{\tilde{f}}}}_{k}} ]}_{w}}{{[ {{{\mathbf{\tilde{f}}}}_{{{k}'}}} ]}_{w}}},\forall k,{k}'$, where ${{\mathbf{\tilde{f}}}_{k}}$ is normalized from ${{\mathbf{f}}_{k}}$ \cite{12}. It should be emphasized that Estimation 2 serves only as a performance baseline, as it relies on knowledge of the local features, which are not available in practice. 
    
    \begin{figure*}[t] \centering
    	\subfigure[Value of surrogate function $G$.] { 
    		{\includegraphics[width=4.8cm]{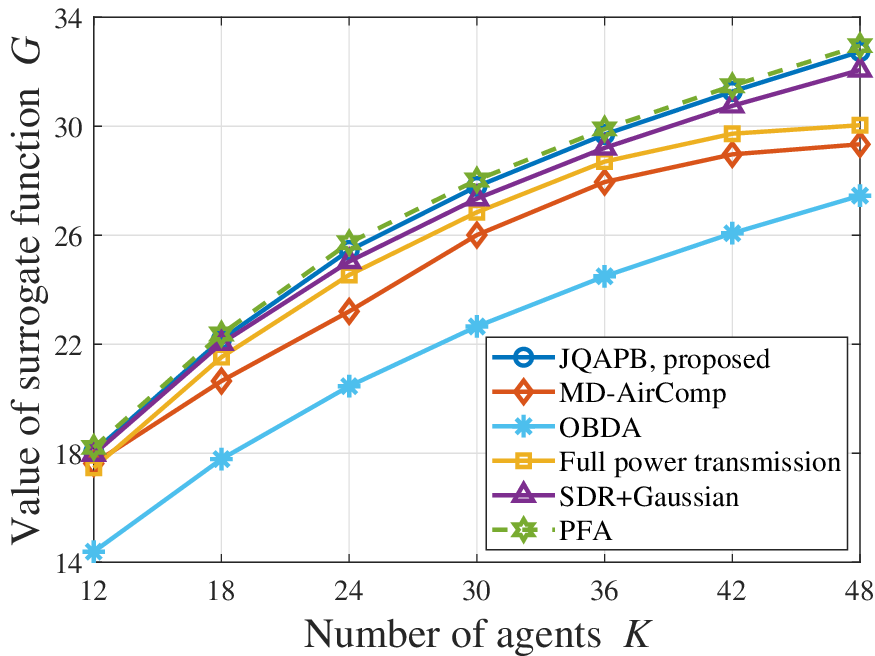}} 
    	}\hspace{0.84cm}     
    	\subfigure[Entropy of posteriors $H$.] {   
    		{\includegraphics[width=4.8cm]{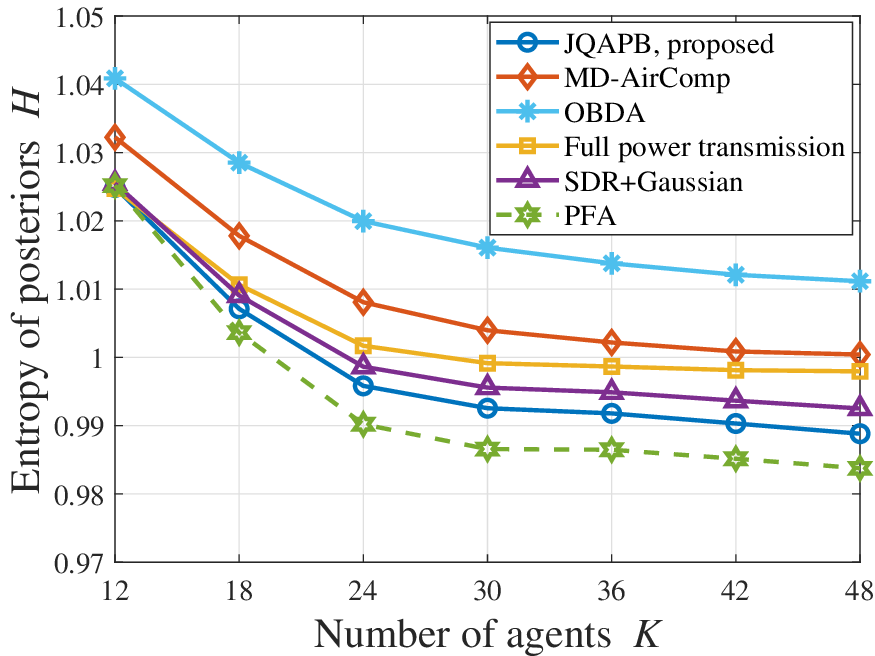}}       
    	}\hspace{0.84cm} 
    	\subfigure[Inference accuracy. ] {   
    		{\includegraphics[width=4.8cm]{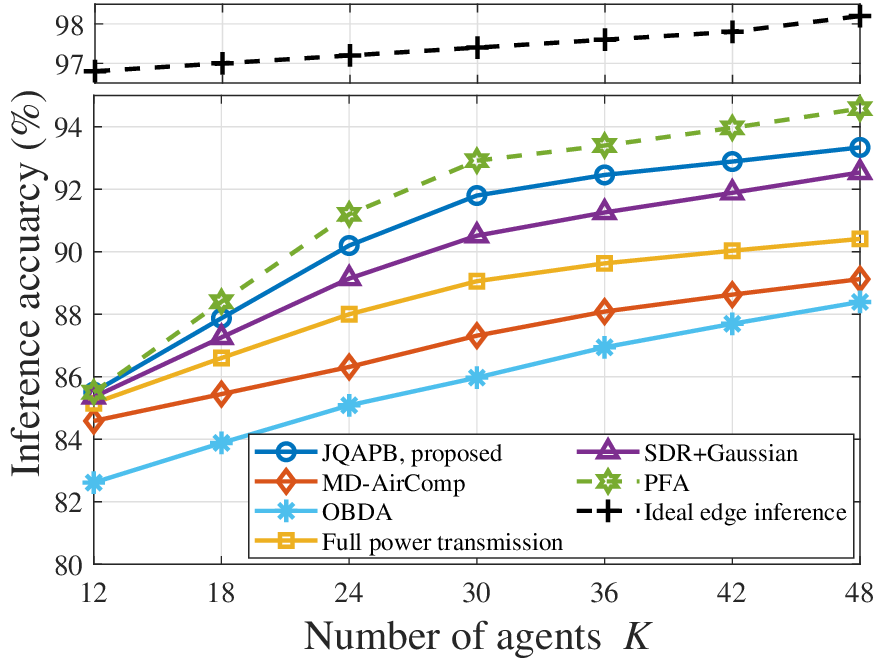}}       
    	}
    	\vspace{3mm}
    	\caption{Different inference performance metrics versus number of agents $K$.}     
    	\label{fig:6}     
    \end{figure*}

    Fig. \ref{fig:4} displays that the normalized aggregation MSE, calculated by $10\lg \left( \frac{\sum\nolimits_{t=1}^{T}{{{\left\| {{{\mathbf{\hat{x}}}}_{t}}-{{\mathbf{x}}_{t}} \right\|}^{2}}}}{\sum\nolimits_{t=1}^{T}{{{\left\| {{\mathbf{x}}_{t}} \right\|}^{2}}}} \right)$, increases with the variance of feature extraction noise $\sigma _{\text{F}}^{2}$, and the minimum MSE is achieved by different values of $\epsilon $ under varying $\sigma _{\text{F}}^{2}$. More specifically, when $\sigma _{\text{F}}^{2}<1$, the local features exhibit strong correlation, and setting $\epsilon =0.6$ yields a close approximation to Estimation 2, resulting in low aggregation MSE. In contrast, Estimation 1 with $\epsilon =0$, which neglects signal correlation, suffers from high MSE in the low-$\sigma _{\text{F}}^{2}$ region and only becomes effective when $\sigma _{\text{F}}^{2}$ is sufficiently large. These observations demonstrate that the proposed beamforming design offers a more general approach compared to \cite{20,21,22,23,24}, as it accommodates varying degrees of feature correlation in edge inference systems. Furthermore, Fig. \ref{fig:4} suggests that Estimation 1 with $\epsilon =0.6$ performs best under our default setting $\sigma _{\text{F}}^{2}=0.5$, which is therefore adopted in the subsequent simulations. 
    
    Fig. \ref{fig:5} showcases the inference accuracy of linear classification under varying values of $J$ and $B$. On the one hand, increasing the available quantization bit $B$ leads to a smaller accuracy gap between \textit{PFA} and \textit{ideal edge inference}. The rational is that a larger $B$ allows each local feature block to be quantized with higher precision, thereby reducing the distortion in the aggregated global features. On the other hand, we observe a performance gap between JQAPB and \textit{PFA}, attributed to the imperfections in over-the-air feature aggregation under practical wireless channels. Interestingly, this gap gradually diminishes as the transmission sequence length $J$ increases, whilst a higher $J$ is required to reduce the gap when $B$ is large. This phenomenon can be explained by that the detection performance of compressive sensing improves with the number of observations (i.e., the length of the received signal ${{\mathbf{y}}_{t}}$), but more observations are needed when the dimension of the signal to be recovered (${{\mathbf{x}}_{t}}$) increases. Nevertheless, increasing $J$ and $B$ leads to higher communication overhead and larger codebook sizes, which should be judiciously balanced against the desired inference accuracy. 
    
    Fig. \ref{fig:sparsity_impact} reveals that as the signal ${\mathbf{x}}_{t}$ becomes less sparse, where sparsity is quantified by the average ratio of zero-valued elements to the signal length, the aggregation MSE increases. This performance degradation arises from the reduced reconstruction capability of the compressive sensing-based SWOMP algorithm. For a fixed signal sparsity, increasing the number of observations $J$ leads to a lower aggregation MSE.
    
    Fig. \ref{fig:6} depicts the curves of the surrogate function $G$, entropy of posterior $H$, and inference accuracy versus number of agents $K$. One can observe that with the increment of $K$, both $G$ and the accuracy rise, while the classification uncertainty represented by $H$ decreases, indicating enhanced inference performance. This improvement is owing to the availability of more diverse viewpoints provided by a larger number of agents, which collectively help suppress feature extraction noise. The consistency between $G$ and accuracy coincides with Theorem 2, corroborating that $G$ is an effective surrogate for characterizing inference performance. Additionally, the proposed JQAPB algorithm outperforms \textit{MD-AirComp}, \textit{OBDA}, \textit{full power transmission}, and \textit{SDR+Gaussian} by 4.48\%, 5.82\%, 2.74\%, and 1.29\% in enhancing the inference accuracy, respectively. This is because that our scheme allocates quantization bits to different feature blocks based on their importance indicators, thereby better preserving local features that contribute more to edge inference. Meanwhile, JQAPB leverages a hybrid RIS to assist feature uploading and adaptively adjusts the active-passive beamforming to align the agent-EN channels, enabling accurate global feature reconstruction in the presence of communication errors. Furthermore, and the inference accuracy of JQAPB is 0.84\% and 6.85\% lower than \textit{PFA} and \textit{ideal edge inference}, respectively, indicating that quantization distortion is the dominant limiting factor under the current setting. Nevertheless, this effect can be mitigated by increasing $B$, as demonstrated in Fig. \ref{fig:5}. 
    
    \subsection{Performance Evaluation for Multi-View Object Recognition}
    
    \begin{figure}[t] \centering
    	\subfigure[Two dimensions $w=1,4$.] { 
    		{\includegraphics[width=4.1cm]{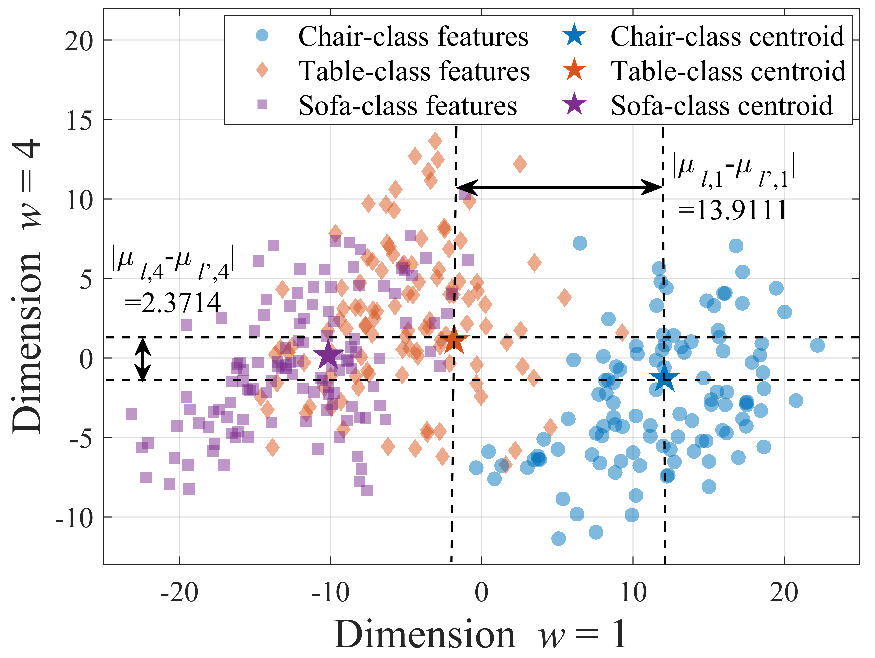}} 
    	}     
    	\subfigure[Three dimensions $w=1,2,4$.] { 
    		{\includegraphics[width=4.1cm]{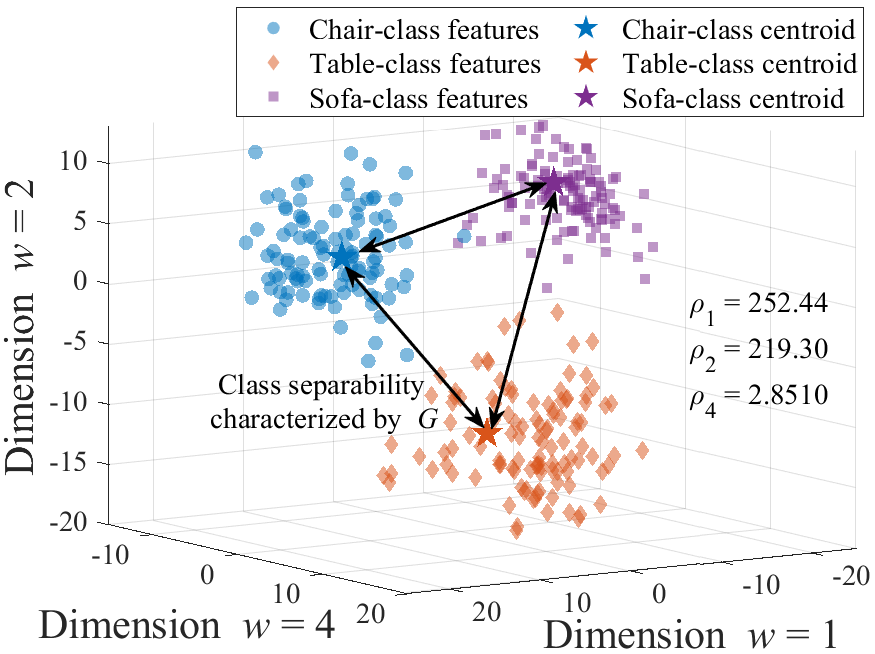}} 
    	}
    	\vspace{0mm}
    	\caption{A visualization of the global features for different classes. }    
    	\label{fig:7}     
    \end{figure}

    This subsection presents the experimental results for multi-view object recognition task. Although the features extracted by MobileNet-v2 (or, more broadly, common DNNs) may not strictly follow a Gaussian mixture distribution, we can exploit the training set feature vectors to fit such a distribution. Fig. \ref{fig:7} visualizes the global features of three representative classes (chair, table and sofa), along with their fitted centroids. Obviously, these classes are more distinguishable in dimensions $w=1,2$ than $w=4$, which aligns with the importance indicator\footnote{There is no theoretical guarantee that $\rho_w$ follows any specific distribution across $w$, as DNN features do not possess an intrinsic ordering. Nevertheless, experimental results indicate that adjacent feature dimensions often exhibit mildly correlated importance, which may stem from shared convolutional filters and correlated channel activations in the backbone network. It is worth emphasizing that our quantization bit allocation does not rely on any predefined feature importance distribution, instead, it may naturally benefit from this empirical phenomenon through a block-wise allocation behavior.} relationship ${{\rho }_{1}}>{{\rho }_{2}}>{{\rho }_{4}}$. Since the derived surrogate function $G$ captures the varying importance across feature dimensions and quantifies the overall class separability in the feature space, enhancing $G$ leads to improved inference accuracy. Consequently, the proposed theoretical framework and JQAPB algorithm are readily applicable to general DNN-enabled inference tasks, whose performance are evaluated in the sequel. 
	
\begin{figure}[t]
    	\begin{tabular}{cc}
    		\begin{minipage}[t]{0.48\linewidth}\centering
    			\includegraphics[width = 4.1cm]{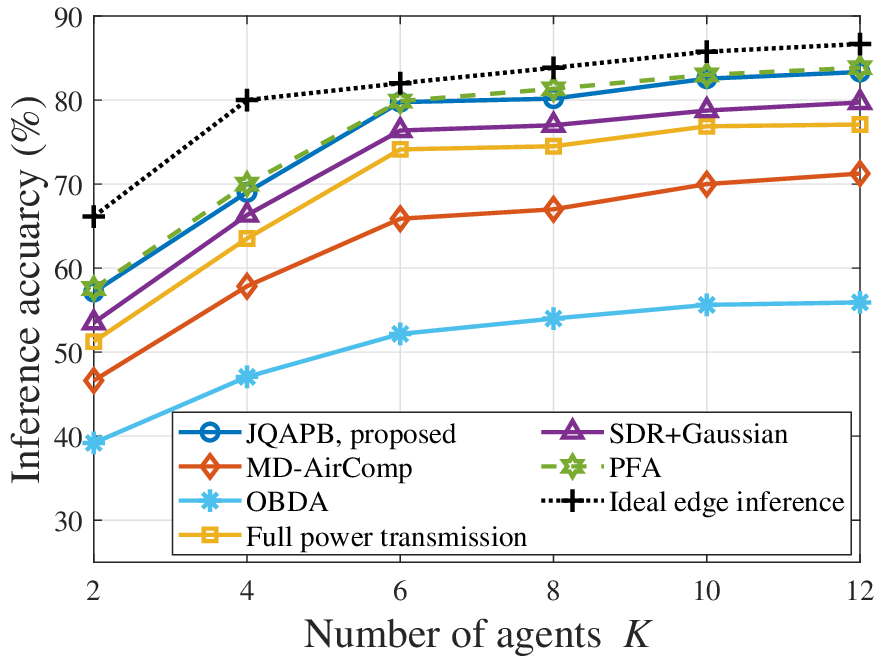}
    			\vspace{1mm}
    			\caption{Inference accuracy versus number of agents $K$.}
    			\label{fig:8}  
    		\end{minipage}
    		\begin{minipage}[t]{0.48\linewidth}\centering
    			\includegraphics[width = 4.1cm]{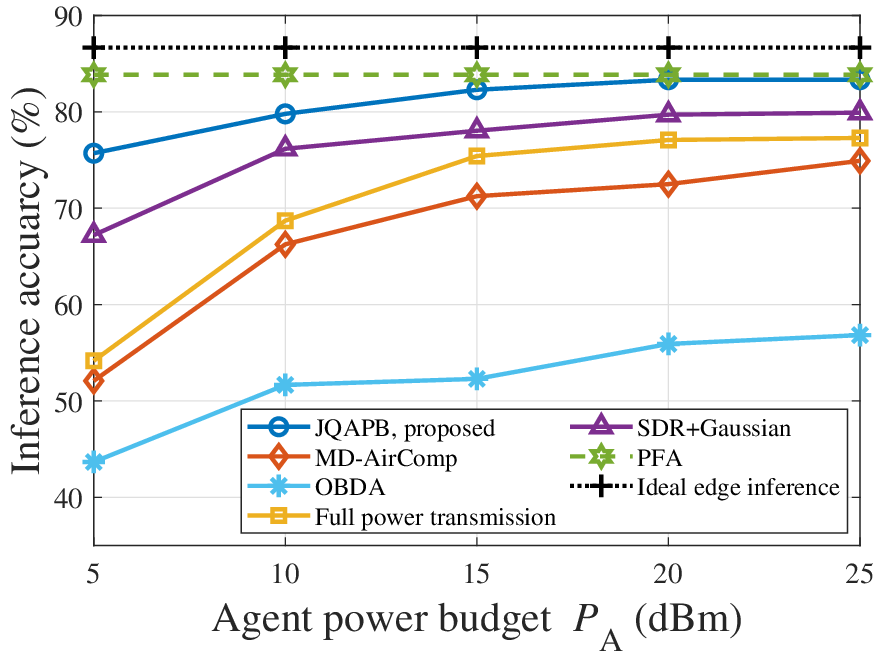}
    			\vspace{1mm}
    			\caption{Inference accuracy versus number of agent power budget $P_\text{A}$.}
    			\label{fig:9} 
    		\end{minipage}
    		\vspace{0mm}	
    	\end{tabular}
    \end{figure}

	Fig. \ref{fig:8} elucidates the inference accuracy of object recognition versus $K$. Similar to the linear classification results in Fig. \ref{fig:6}, the accuracy increases with $K$, and the proposed JQAPB approach outperforms most benchmark methods. Specifically, JQAPB achieves inference accuracy that is 12.23\%, 24.67\%, 5.77\%, and 3.39\% higher than those of \textit{MD-AirComp}, \textit{OBDA}, \textit{full power transmission}, and \textit{SDR+Gaussian}, respectively. Meanwhile, its accuracy is only 0.51\% and 3.33\% lower than \textit{PFA} and \textit{ideal edge inference}. Notably, the performance gap caused by feature quantization is smaller than that observed in linear classification. This suggests that in DNN-based inference, a small number of important features tends to dominate the inference outcome, thus mitigating the impact of quantization.
    
    The curves of inference accuracy versus agent power budget ${{P}_{\text{A}}}$ is plotted in Fig. \ref{fig:9}. The proposed JQAPB is capable of flexibly adjusting the agent transmission coefficients, so as to compensate for the misalignment error during feature aggregation. As a result, it effectively harnesses the performance gains brought by increased power budget, achieving inference accuracy close to that of \textit{PFA} and \textit{ideal edge inference}, whilst outperforming other benchmark methods. 
    
    \begin{figure}[t] \centering
    	\subfigure[Inference accuracy versus total number of elements $N$.] { 
    		{\includegraphics[width=4.1cm]{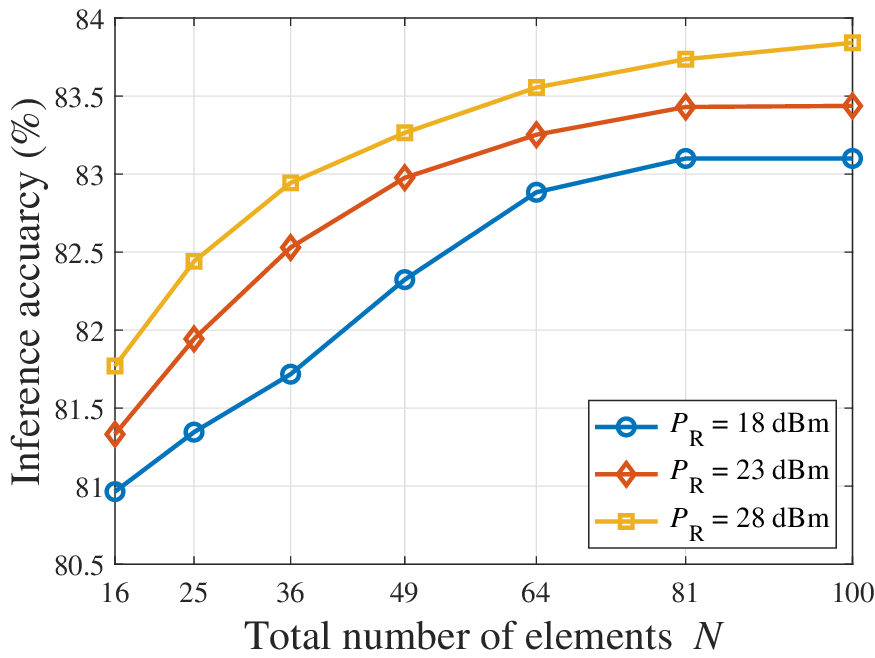}} 
    	}     
    	\subfigure[Inference accuracy versus number of active elements $N_\text{a}$.] { 
    		{\includegraphics[width=4.1cm]{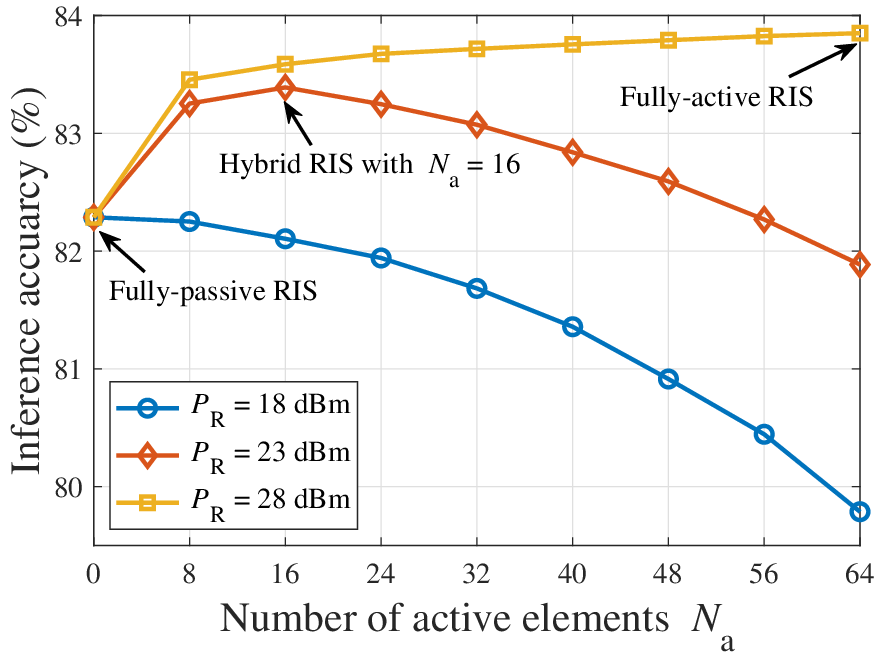}} 
    	}
    	\vspace{0mm}
    	\caption{Inference accuracy under different hybrid RIS configurations.}     
    	\label{fig:10}     
    \end{figure}
    
    Fig. \ref{fig:10} examines the impact of different hybrid RIS configurations on inference accuracy. In Fig. \ref{fig:10} (a), we vary the total number of RIS elements $N$, while setting the number of active elements as ${{N}_{\text{a}}}=\left\lfloor {N}/{8}\right\rfloor $. The results show that the inference accuracy increases with $N$, owing to the enhanced spatial DoFs for improving the feature uploading channels. However, the accuracy saturates when $N$ becomes sufficiently large, as the limited amplification power of the RIS ${{P}_{\text{R}}}$ emerges as the performance bottleneck. 
    
    In Fig. \ref{fig:10} (b), we fix $N=\text{64}$ and change ${{N}_{\text{a}}}$, where ${{N}_{\text{a}}}=\text{0}$ and 64 correspond to a fully-passive RIS and a fully-active RIS, respectively. When ${{P}_{\text{R}}}=\text{18}\text{ dBm}$, the inference accuracy monotonically decreases with ${{N}_{\text{a}}}$, indicating that the fully-passive RIS yields the best performance in low power regine. This is because the square-order beamforming gain provided by the passive elements dominates the contribution of power amplification. In contrast, fully-active RIS realizes the highest accuracy when ${{P}_{\text{R}}}=\text{28}\text{ dBm}$, as the ample power budget allows the active elements to fully leverage their amplification capability. Furthermore, at a moderate power budget of ${{P}_{\text{R}}}=\text{23}\text{ dBm}$, hybrid RIS with ${{N}_{\text{a}}}=\text{16}$ strikes a good balance between passive beamforming and active power amplification, thereby outperforming other types of RISs. In conclusion, the proposed JQAPB offers a generalized RIS reflection beamforming approach that can flexibly adapt to various RIS configurations. 
    
    \begin{figure}[t]\centering
    	\includegraphics[width = 4.8cm]{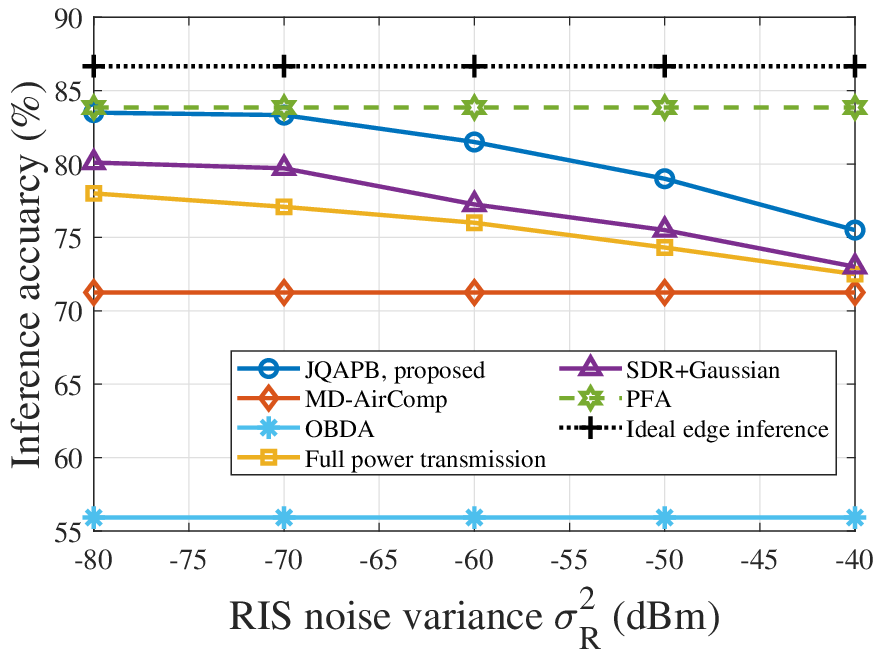}
    	\vspace{1mm}
    	\caption{Inference accuracy versus RIS noise variance $\sigma^2_{\text{R}}$.}
    	\label{fig:RIS_noise_impact} 
    \end{figure}
    
   Fig. \ref{fig:RIS_noise_impact} illustrates the impact of the RIS noise variance $\sigma^2_{\mathrm{R}}$ on inference accuracy. As $\sigma^2_{\mathrm{R}}$ increases, more severe channel noise is introduced into the over-the-air feature aggregation, thereby degrading the inference performance. Nevertheless, owing to the carefully designed active-passive reflection beamforming at the hybrid RIS, our JQAPB consistently outperforms baseline methods by 3.45\%$\sim$24.65\%.
    
	\section{Conclusion} \label{sec:conclusion}	
	
	In this work, we proposed the HRD-AirComp scheme to enable communication-efficient feature aggregation in multi-agent collaborative edge inference systems. A key advantage of the proposed scheme lies in its compatibility with existing digital communication infrastructures, achieved by employing vector quantization to map each feature block into a discrete codeword, which is then used to select a corresponding transmission sequence from the modulation codebook. In addition, HRD-AirComp incorporates task-oriented feature quantization and active-passive beamforming design. Specifically, based on the derived closed-form surrogate function for inference accuracy, the JQAPB algorithm is developed to jointly optimize the quantization bit allocation, AirComp transceivers at the agents and EN, as well as the hybrid RIS reflection beamforming. Our theoretical analysis and numerical results reveal that: 1) HRD-AirComp realizes effective preservation of critical features during transmission by quantizing feature blocks with differentiated codebooks and adaptively allocated bits. 2) The deployment of hybrid RIS enhances inference performance, as it proactively reconfigures the over-the-air feature aggregation process through simultaneous signal amplification and reflection. 3) Thanks to the closed-form updates of optimization variables, JQAPB achieves significantly higher implementation efficiency compared to traditional SDR-based methods. 4) In real-world object recognition tasks, the proposed scheme improves inference accuracy by 12.23\% to 24.67\% over existing baselines, with a performance gap of only 0.51\% relative to the idealized \textit{PFA} case. As HRD-AirComp provides a general framework for over-the-air data computation within digital communication systems, future work will explore its broader applicability to other edge AI scenarios, such as federated learning/fine-tuning and distributed consensus. 
	
	\begin{appendices} 
		\section{Proof of Theorem 1}
		
		The aggregation MSE of the $t$-th feature block is expressed as 
		\begin{align}
			&\mathbb{E}\left[ {{\left\| {{{\mathbf{\hat{f}}}}_{t}}-{{\mathbf{f}}_{t}} \right\|}^{2}} \right]=\frac{1}{{{K}^{2}}}\mathbb{E}\left[ {{\left\| {{\mathbf{Q}}_{t}}{{{\mathbf{\hat{x}}}}_{t}}-\sum\limits_{k\in \mathcal{K}}{{{\mathbf{f}}_{k,t}}} \right\|}^{2}} \right]\nonumber\\
			&=\frac{1}{{{K}^{2}}}\mathbb{E}\left[ {{\left\| {{\mathbf{Q}}_{t}}{{{\mathbf{\hat{x}}}}_{t}}-{{\mathbf{Q}}_{t}}{{\mathbf{x}}_{t}}+{{\mathbf{Q}}_{t}}{{\mathbf{x}}_{t}}-\sum\limits_{k\in \mathcal{K}}{{{\mathbf{f}}_{k,t}}} \right\|}^{2}} \right] \nonumber\\
			&\overset{\left( a \right)}{\mathop{\le }}\!\frac{2}{{{K}^{2}}}\underbrace{\mathbb{E}\left[ {{\left\| {{\mathbf{Q}}_{t}}\left( {{{\mathbf{\hat{x}}}}_{t}}-{{\mathbf{x}}_{t}} \right) \right\|}^{2}} \right]}_{{{A}_{1}}}+\frac{2}{{{K}^{2}}}\underbrace{\mathbb{E}\left[ {{\left\| {{\mathbf{Q}}_{t}}{{\mathbf{x}}_{t}}-\sum\limits_{k\in \mathcal{K}}{{{\mathbf{f}}_{k,t}}} \right\|}^{2}} \right]}_{{{A}_{2}}}, \label{eq:31}
		\end{align}
		where $\left( a \right)$ follows inequality ${{\left\| \mathbf{a}+\mathbf{c} \right\|}^{2}}\le 2\left( {{\left\| \mathbf{a} \right\|}^{2}}+{{\left\| \mathbf{c} \right\|}^{2}} \right)$. 
		
		To proceed with, we bound the term ${{A}_{1}}$ below. 
		\begin{align}
			{{A}_{1}}\overset{\left( a \right)}{\mathop{\le }}\,{{\left\| {{\mathbf{Q}}_{t}} \right\|}_{2}}\mathbb{E}\left[ {{\left\| {{{\mathbf{\hat{x}}}}_{t}}-{{\mathbf{x}}_{t}} \right\|}^{2}} \right]\overset{\left( b \right)}{\mathop{\le }}\,{{2}^{{{B}_{t}}}}\mathbb{E}\left[ {{\left\| {{{\mathbf{\hat{x}}}}_{t}}-{{\mathbf{x}}_{t}} \right\|}^{2}} \right], \label{eq:32}
		\end{align}
		where $\left( a \right)$ and $\left( b \right)$ hold by $\left\| \mathbf{Ac} \right\|\le {{\left\| \mathbf{A} \right\|}_{2}}\left\| \mathbf{c} \right\|\le {{\left\| \mathbf{A} \right\|}_{F}}\left\| \mathbf{c} \right\|$, with ${{\left\| \cdot  \right\|}_{2}}$ and ${{\left\| \cdot  \right\|}_{F}}$ being the spectral norm and Frobenius norm, respectively. Besides, ${{\left\| \mathbf{Q} \right\|}_{F}}={{I}_{t}}={{2}^{{{B}_{t}}}}$ since each codeword in the Grassmannian codebook is a unit vector. According to Assumption 2, we expand the expectation term in (\ref{eq:32}) as 
		\begin{align}
			&\!\!\!\!{{A}_{1}}\le {{2}^{{{B}_{t}}}}\eta \mathbb{E}\left[ {{\left\| {{\mathbf{y}}_{t}}-{{\mathbf{s}}_{t}} \right\|}^{2}} \right]\nonumber\\
			&\!\!\!={{2}^{{{B}_{t}}}}\eta \mathbb{E}\!\left[ {{\left\| \sum\limits_{k\in \mathcal{K}}{\left( {{\nu }_{k}}\mathbf{h}_{k}^{\text{T}}\mathbf{b}\!-\! 1 \right){{\mathbf{s}}_{k,t}}}\!+\!{{\mathbf{Z}}_{\text{R},t}}{{\mathbf{\Phi }}_{\text{a}}}\mathbf{H}_{\text{RE}}^{\text{T}}\mathbf{b}\!+\!{{\mathbf{Z}}_{\text{E},t}}\mathbf{b} \right\|}^{2}} \right] \nonumber\\ 
			&={{2}^{{{B}_{t}}}}\eta \mathbb{E}\bigg[ {{\left\| \sum\nolimits_{k\in \mathcal{K}}{\left( {{\nu }_{k}}\mathbf{h}_{k}^{\text{T}}\mathbf{b}\!-\! 1 \right){{\mathbf{s}}_{k,t}}} \right\|}^{2}}\!+\!{{\left\| {{\mathbf{Z}}_{\text{R},t}}{{\mathbf{\Phi }}_{\text{a}}}\mathbf{H}_{\text{RE}}^{\text{T}}\mathbf{b} \right\|}^{2}}\nonumber\\
			&\qquad\qquad\qquad\quad\qquad\qquad\qquad\qquad\ \ +{{\left\| {{\mathbf{Z}}_{\text{E},t}}\mathbf{b} \right\|}^{2}} \bigg]  \nonumber\\ 
			&={{2}^{{{B}_{t}}}}\eta \bigg\{ \mathbb{E}\!\left[ {{\left\| \sum\nolimits_{k\in \mathcal{K}}\!\!{\left( {{\nu }_{k}}\mathbf{h}_{k}^{\text{T}}\mathbf{b}\!-\!1 \right){{\mathbf{s}}_{k,t}}} \right\|}^{2}} \right]\!\!+\!\sigma _{\text{R}}^{2}{{\left\| {{\mathbf{\Phi }}_{\text{a}}}\mathbf{H}_{\text{RE}}^{\text{T}}\mathbf{b} \right\|}^{2}}\nonumber\\
			&\qquad\qquad\qquad\quad\qquad\qquad\qquad\qquad\quad\ +\sigma _{\text{E}}^{2}{{\left\| \mathbf{b} \right\|}^{2}} \bigg\} \nonumber\\ 
			&={{2}^{{{B}_{t}}}}\eta \bigg\{ \!\sum\limits_{k\in \mathcal{K}}{\sum\limits_{{k}'\in \mathcal{K}}\!\!{\mathbb{E}\!\left[ {{\left( {{\nu }_{k}}\mathbf{h}_{k}^{\text{T}}\mathbf{b}\!-\! 1 \right)}^{*}}\left( {{\nu }_{{{k}'}}}\mathbf{h}_{{{k}'}}^{\text{T}}\mathbf{b}\!-\! 1 \right)\mathbf{s}_{k,t}^{\text{H}}{{\mathbf{s}}_{{k}',t}} \right]}}\nonumber\\
			&\qquad\qquad\qquad\qquad\quad+\sigma _{\text{R}}^{2}{{\left\| {{\mathbf{\Phi }}_{\text{a}}}\mathbf{H}_{\text{RE}}^{\text{T}}\mathbf{b} \right\|}^{2}}+\sigma _{\text{E}}^{2}{{\left\| \mathbf{b} \right\|}^{2}} \bigg\} \nonumber\\ 
			&\overset{\left( a \right)}{\mathop{\le }}\,{{2}^{{{B}_{t}}}}\eta \bigg\{ {{\beta }^{2}}J\sum\limits_{k\in \mathcal{K}}{\sum\limits_{{k}'\in \mathcal{K}}{{{u}_{k,{k}'}}{{\left( {{\nu }_{k}}\mathbf{h}_{k}^{\text{T}}\mathbf{b}-1 \right)}^{*}}\left( {{\nu }_{{{k}'}}}\mathbf{h}_{{{k}'}}^{\text{T}}\mathbf{b}-1 \right)}}\nonumber\\
			&\qquad\qquad\qquad\qquad\quad+\sigma _{\text{R}}^{2}{{\left\| {{\mathbf{\Phi }}_{\text{a}}}\mathbf{H}_{\text{RE}}^{\text{T}}\mathbf{b} \right\|}^{2}}+\sigma _{\text{E}}^{2}{{\left\| \mathbf{b} \right\|}^{2}} \bigg\},
			\label{eq:33}
		\end{align}
		where $\left( a \right)$ is due to Assumption 3-4, which implies $\mathbb{E}\left[ \mathbf{s}_{k,t}^{\text{H}}{{\mathbf{s}}_{{k}',t}} \right]=\mathbb{E}\left[ {{\beta }_{k,t}}{{\beta }_{{k}',t}} \right]J{{u}_{k,{k}'}}\le {{\beta }^{2}}J{{u}_{k,{k}'}}$. 
		
		Next, the term ${{A}_{2}}$ is bounded as follows
		\begin{align}
			&{{A}_{2}}\!=\!\mathbb{E}\left[ {{\left\| \sum\nolimits_{k\in \mathcal{K}}{{{\beta }_{k,t}}{{\mathbf{Q}}_{t}}{{\mathbf{x}}_{k,t}}}\!-\!\sum\nolimits_{k\in \mathcal{K}}{{{\beta }_{k,t}}{{\mathbf{v}}_{k,t}}} \right\|}^{2}} \right]\nonumber\\
			&\overset{\left( a \right)}{\mathop{=}}K\mathbb{E}\left[ {{\left\| {{\beta }_{k,t}}\left( {{\mathbf{Q}}_{t}}{{\mathbf{x}}_{k,t}}\!-\!{{\mathbf{v}}_{k,t}} \right) \right\|}^{2}} \right]\!\overset{\left( b \right)}{\mathop{\le }}\! K{{\beta }^{2}}\mathbb{E}\left[ {{\left\| {{\mathbf{Q}}_{t}}{{\mathbf{x}}_{k,t}}\!-\!{{\mathbf{v}}_{k,t}} \right\|}^{2}} \right],  \label{eq:34}
		\end{align}
		where $\left( a \right)$ is because the quantization errors of different agents are dependent, $\left( b \right)$ follows Assumption 4. Then, we introduce the following lemma to characterize the average distortion for block normalization. 
		
		\textit{Lemma 1:} For the Grassmannian codebook ${{\mathbf{Q}}_{t}}$ with ${{I}_{t}}={{2}^{{{B}_{t}}}}$ uniformly distributed codewords and dimension $D$, the expected error for quantizing block normalization ${{\mathbf{v}}_{k,t}}$ is bounded by [27]
		\begin{align}
			\mathbb{E}\left[ {{\left\| {{\mathbf{Q}}_{t}}{{\mathbf{x}}_{k,t}}-{{\mathbf{v}}_{k,t}} \right\|}^{2}} \right]\le {{2}^{-\frac{2{{B}_{t}}}{D-1}}}. \label{eq:35}
		\end{align}
		
		Accordingly, we have ${{A}_{2}}\le K{{\beta }^{2}}{{2}^{-\frac{2{{B}_{t}}}{D-1}}}$. By substituting this as well as (\ref{eq:33}) into (\ref{eq:31}), we can derive the upper bound of the mean square feature aggregation error in Theorem 1. This completes the proof.
		
		\section{Proof of Theorem 2}
		
		Given the entropy of posteriors $H$ in (1), we can further derive it as
		\begin{align}
			H&\overset{\left( a \right)}{\mathop{=}}\int{-\Pr ( {\mathbf{\hat{f}}} )\sum\limits_{l=1}^{L}{\frac{\Pr ( \mathbf{\hat{f}}|l )}{L\Pr ( {\mathbf{\hat{f}}} )}\ln \frac{\Pr ( \mathbf{\hat{f}}|l )}{L\Pr ( {\mathbf{\hat{f}}} )}}\text{d}\mathbf{\hat{f}}}\nonumber\\
			&=\frac{1}{L}\sum\limits_{l=1}^{L}{\int{\Pr ( \mathbf{\hat{f}}|l )\ln \frac{L\Pr ( {\mathbf{\hat{f}}} )}{\Pr ( \mathbf{\hat{f}}|l )}\text{d}\mathbf{\hat{f}}}}\nonumber\\
			&\overset{\left( b \right)}{\mathop{=}}\,\frac{1}{L}\sum\limits_{l=1}^{L}{\int{\Pr ( \mathbf{\hat{f}}|l )\ln \frac{\sum\nolimits_{{l}'=1}^{L}{\Pr ( \mathbf{\hat{f}}|{l}' )}}{\Pr ( \mathbf{\hat{f}}|l )}\text{d}\mathbf{\hat{f}}}},\label{eq:36}
		\end{align}
		where $\left( a \right)$ leverages the Bayes formula to express $\Pr ( l|\mathbf{\hat{f}} )$ as ${\Pr ( \mathbf{\hat{f}}|l )}/{( L\Pr ( {\mathbf{\hat{f}}} ) )}$ with $\Pr \left( l \right)=\frac{1}{L}$. $\left( b \right)$ follows the formula of total probability, i.e., $\Pr ( {\mathbf{\hat{f}}} )=\sum\nolimits_{{l}'=1}^{L}{\Pr ( \mathbf{\hat{f}}|{l}' )\Pr \left( {{l}'} \right)}=\frac{1}{L}\sum\nolimits_{{l}'=1}^{L}{\Pr ( \mathbf{\hat{f}}|{l}' )}$. According to Corollary 1, we have $\Pr ( \mathbf{\hat{f}}|l )=\frac{1}{{{\left( 2\pi  \right)}^{\frac{W}{2}}}{{\left| {\mathbf{\hat{C}}} \right|}^{\frac{1}{2}}}}\exp \{ -\frac{1}{2}{{( \mathbf{\hat{f}}-{{{\hat{\bm{\mu} }}}_{l}} )}^{\text{T}}}{{{\mathbf{\hat{C}}}}^{-1}}( \mathbf{\hat{f}}-{{{\hat{\bm{\mu} }}}_{l}} ) \}$. By modifying the integration variable into ${{\mathbf{g}}_{l}}=\mathbf{\hat{f}}-{{\hat{\bm{\mu} }}_{l}}$, we rewrite $H$ in (\ref{eq:37}),
		\begin{figure*}
			\begin{align}
				H&=\frac{1}{L}\sum\limits_{l=1}^{L}{\int{\Pr \left( {{\mathbf{g}}_{l}}|l \right)\ln \frac{\sum\nolimits_{{l}'=1}^{L}{\exp \left\{ -\frac{1}{2}{{\left( {{\mathbf{g}}_{l}}+{{{\hat{\bm{\mu} }}}_{l}}-{{{\hat{\bm{\mu} }}}_{{{l}'}}} \right)}^{\text{T}}}{{{\mathbf{\hat{C}}}}^{-1}}\left( {{\mathbf{g}}_{l}}+\hat{\bm{\mu} }-{{{\hat{\bm{\mu} }}}_{{{l}'}}} \right) \right\}}}{\exp \left\{ -\frac{1}{2}{{\mathbf{g}}_{l}}^{\text{T}}{{{\mathbf{\hat{C}}}}^{-1}}{{\mathbf{g}}_{l}} \right\}}\text{d}{{\mathbf{g}}_{l}}}}\nonumber\\
				&=\frac{1}{L}\sum\limits_{l=1}^{L}{\int{\Pr \left( {{\mathbf{g}}_{l}}|l \right)\ln \left[ \sum\limits_{{l}'=1}^{L}{\exp \left\{ -{{\left( {{{\hat{\bm{\mu} }}}_{l}}-{{{\hat{\bm{\mu} }}}_{{{l}'}}} \right)}^{\text{T}}}{{{\mathbf{\hat{C}}}}^{-1}}{{\mathbf{g}}_{l}} \right\}\cdot \exp \left\{ -\frac{1}{2}{{\left( {{{\hat{\bm{\mu} }}}_{l}}-{{{\hat{\bm{\mu} }}}_{{{l}'}}} \right)}^{\text{T}}}{{{\mathbf{\hat{C}}}}^{-1}}\left( {{{\hat{\bm{\mu} }}}_{l}}-{{{\hat{\bm{\mu} }}}_{{{l}'}}} \right) \right\}} \right]\text{d}{{\mathbf{g}}_{l}}}}\nonumber\\
				&\overset{\left( a \right)}{\mathop{\ge }}\,\frac{1}{L}\sum\limits_{l=1}^{L}{\ln \left[ \sum\limits_{{l}'=1}^{L}{\exp \left\{ -{{\left( {{{\hat{\bm{\mu} }}}_{l}}-{{{\hat{\bm{\mu} }}}_{{{l}'}}} \right)}^{\text{T}}}{{{\mathbf{\hat{C}}}}^{-1}}\int{\Pr \left( {{\mathbf{g}}_{l}}|l \right){{\mathbf{g}}_{l}}\text{d}{{\mathbf{g}}_{l}}} \right\}\cdot \exp \left\{ -\frac{1}{2}{{\left( {{{\hat{\bm{\mu} }}}_{l}}-{{{\hat{\bm{\mu} }}}_{{{l}'}}} \right)}^{\text{T}}}{{{\mathbf{\hat{C}}}}^{-1}}\left( {{{\hat{\bm{\mu} }}}_{l}}-{{{\hat{\bm{\mu} }}}_{{{l}'}}} \right) \right\}} \right]}\nonumber\\
				&\overset{\left( b \right)}{\mathop{=}}\,\frac{1}{L}\sum\limits_{l=1}^{L}{\ln \left[ 1+\sum\limits_{{l}'=1,{l}'\ne l}^{L}{\exp \left\{ -\frac{1}{2}{{\left( {{{\hat{\bm{\mu} }}}_{l}}-{{{\hat{\bm{\mu} }}}_{{{l}'}}} \right)}^{\text{T}}}{{{\mathbf{\hat{C}}}}^{-1}}\left( {{{\hat{\bm{\mu} }}}_{l}}-{{{\hat{\bm{\mu} }}}_{{{l}'}}} \right) \right\}} \right]}\triangleq {{H}^{\text{low}}}, \label{eq:37}
			\end{align}
		\end{figure*}
		where $\left( a \right)$ is owing to the Jensen’s inequality, since the log-sum-exp function in the integration is convex w.r.t. ${{\mathbf{g}}_{l}}$, thus a lower bound of $H$ can be obtained. $\left( b \right)$ is due to $\int{\Pr \left( {{\mathbf{g}}_{l}}|l \right){{\mathbf{g}}_{l}}\text{d}{{\mathbf{g}}_{l}}}=\mathbb{E}\left( {{\mathbf{g}}_{l}} \right)=\mathbb{E}( {\mathbf{\hat{f}}}|l)-{{\hat{\bm{\mu} }}_{l}}=0$. 
		
		Subsequently, we define ${{G}_{l,{l}'}}={{\left( {{{\hat{\bm{\mu} }}}_{l}}-{{{\hat{\bm{\mu} }}}_{{{l}'}}} \right)}^{\text{T}}}{{\mathbf{\hat{C}}}^{-1}}\left( {{{\hat{\bm{\mu} }}}_{l}}-{{{\hat{\bm{\mu} }}}_{{{l}'}}} \right)$, and $G=\frac{1}{L\left( L-1 \right)}\sum\nolimits_{l<{l}'\le L}{{{G}_{l,{l}'}}}$ denotes the average version for all class pairs $\forall \left( l,{l}' \right)$. One can observe that ${{H}^{\text{low}}}$ is a function of $\left\{ {{G}_{l,{l}'}} \right\}$, we further approximate it using the first-order Talor expansion at $\left\{ G \right\}$ in the sequel
		\begin{align}
			{{H}^{\text{low}}}&=\frac{1}{L}\sum\limits_{l=1}^{L}{\ln \left[ 1+\sum\limits_{{l}'=1,{l}'\ne l}^{L}{\exp \left\{ -\frac{{{G}_{l,{l}'}}}{2} \right\}} \right]}\nonumber\\
			&\approx \frac{1}{L}\sum\limits_{l=1}^{L}{\ln \left[ 1+\sum\limits_{{l}'=1,{l}'\ne l}^{L}{\exp \left\{ -\frac{G}{2} \right\}} \right]}\nonumber\\
			&\ \ \ +\sum\limits_{l<{l}'\le L}{{{\left. \frac{\partial H}{\partial {{G}_{l,{l}'}}} \right|}_{{{G}_{l,{l}'}}=G}}\left( {{G}_{l,{l}'}}-G \right)},\label{eq:38}
		\end{align}
		where ${{\left. \frac{\partial H}{\partial {{G}_{l,{l}'}}} \right|}_{{{G}_{l,{l}'}}=G}}=-\frac{1}{2L}\frac{\exp \left\{ {-G}/{2}\; \right\}}{1+\left( L-1 \right)\exp \left\{ {-G}/{2} \right\}}=c_0$ is a constant, then the second term $c_0\sum\nolimits_{l<{l}'\le L}{\left( {{G}_{l,{l}'}}-G \right)}=c_0\left[ \sum\nolimits_{l<{l}'\le L}{{{G}_{l,{l}'}}}-L\left( L-1 \right)G \right]=0$. As a result, ${{H}^{\text{low}}}$ can be approximated by
		\begin{align}
			{{H}^{\text{low}}}\approx \frac{1}{L}\sum\limits_{l=1}^{L}{\ln \left[ 1+\left( L-1 \right){{\text{e}}^{{-G}/{2}\;}} \right]}. \label{eq:39}
		\end{align}
		This completes the proof. 
		
		\section{Proof of Theorem 3}
		To verify the convergence of \textbf{Algorithm 1}, we need to prove that both inner and outer iteration can be converged within a finite number of steps. Define the objective value of \textbf{P1} in (\ref{eq:14a}) as $f(\cdot)$. For the inner iteration, we derive an upper bound in (\ref{eq:26}), and the corresponding objective value is denoted as $\tilde{f}(\cdot)$. Given $\bar{\bm{\alpha}}$, $\bar{{\mathbf{q}}}_{{{\mathbf{\Lambda }}_{\text{a}}}}$, and $\bar{\mathbf{d}}$ from the previous iteration, we have
		\begin{align}
			f(\bar{\bm{\alpha}},\bar{{\mathbf{q}}}_{{{\mathbf{\Lambda }}_{\text{a}}}},\bar{\mathbf{d}}) \overset{\left( a \right)}{\mathop{= }} \tilde{f}(\bar{\bm{\alpha}},\bar{{\mathbf{q}}}_{{{\mathbf{\Lambda }}_{\text{a}}}},\bar{\mathbf{d}})\overset{\left( b \right)}{\mathop{\le }}\tilde{f}(\bm{\alpha},{\mathbf{q}}_{{{\mathbf{\Lambda }}_{\text{a}}}},\mathbf{d}) \overset{\left( c \right)}{\mathop{\le }} f(\bm{\alpha},{\mathbf{q}}_{{{\mathbf{\Lambda }}_{\text{a}}}},\mathbf{d}), \label{eq:40}
		\end{align}
		where $(a)$ holds since the approximation in (\ref{eq:26}) is tight with given point. $(b)$ holds owing to that the variables related to hybrid RIS reflection beamforming ${\mathbf{\Phi }}$ are optimally updated through the closed-form expressions (\ref{eq:27}), (\ref{eq:29}), and (\ref{eq:25}). $(c)$ follows that $\tilde{f}$ is a lower bound of $f$ as optimizing the right-hand-side of (\ref{eq:26}) is more strict.
			
		For the outer iteration, we denote $\hat{f}(\cdot)$ as the objective value in (\ref{eq:16a}) after Taylor expansion, then the following inequalities can be derived:
		\begin{align}
			&f(\left\{ {\bar{B}_{t}} \right\}_{t=1}^{T},\left\{ {\bar{\nu }_{k}} \right\}_{k=1}^{K},\bar{\mathbf{b}},\bar{\mathbf{\Phi }}) = \hat{f}(\left\{ {\bar{B}_{t}} \right\}_{t=1}^{T},\left\{ {\bar{\nu }_{k}} \right\}_{k=1}^{K},\bar{\mathbf{b}},\bar{\mathbf{\Phi }})\nonumber\\
			&\overset{\left( a \right)}{\mathop{\le }}\hat{f}(\left\{ {{B}_{t}} \right\}_{t=1}^{T}\!,\left\{ {\bar{\nu }_{k}} \right\}_{k=1}^{K},\bar{\mathbf{b}},\bar{\mathbf{\Phi }})\!\le\! f(\left\{ {{B}_{t}} \right\}_{t=1}^{T}\!,\left\{ {\bar{\nu }_{k}} \right\}_{k=1}^{K},\bar{\mathbf{b}},\bar{\mathbf{\Phi }})\nonumber\\
			&\overset{\left( b \right)}{\mathop{\le }}f(\left\{ {{B}_{t}} \right\}_{t=1}^{T}\!,\left\{ {{\nu }_{k}} \right\}_{k=1}^{K},{\mathbf{b}},\bar{\mathbf{\Phi }})\overset{\left( c \right)}{\mathop{\le }}f(\left\{ {{B}_{t}} \right\}_{t=1}^{T}\!,\left\{ {{\nu }_{k}} \right\}_{k=1}^{K},{\mathbf{b}},{\mathbf{\Phi }}), \label{eq:41}
		\end{align}
		where $(a)$ is attributed to that \textbf{SP1} is addressed optimally with solution $\left\{ {{B}_{t}} \right\}_{t=1}^{T}$. $(b)$ is obtained via the optimal solution of convex optimization subproblems \textbf{SP2} and \textbf{SP3}. $(c)$ follows the operation of inner iteration in (\ref{eq:40}). (\ref{eq:41}) indicates that $f$ is monotonically non-decreasing after each iteration. Besides, it is obvious that $f$ is upper-bounded due to the limited power budget and quantization bit, thus \textbf{Algorithm 1} is guaranteed to converge to a stationary solution to \textbf{P1}. 
		
	\end{appendices}

	\bibliographystyle{IEEEtran}
	\bibliography{reference}

\begin{thebibliography}{10}
\providecommand{\url}[1]{#1}
\csname url@samestyle\endcsname
\providecommand{\newblock}{\relax}
\providecommand{\bibinfo}[2]{#2}
\providecommand{\BIBentrySTDinterwordspacing}{\spaceskip=0pt\relax}
\providecommand{\BIBentryALTinterwordstretchfactor}{4}
\providecommand{\BIBentryALTinterwordspacing}{\spaceskip=\fontdimen2\font plus
\BIBentryALTinterwordstretchfactor\fontdimen3\font minus
  \fontdimen4\font\relax}
\providecommand{\BIBforeignlanguage}[2]{{%
\expandafter\ifx\csname l@#1\endcsname\relax
\typeout{** WARNING: IEEEtran.bst: No hyphenation pattern has been}%
\typeout{** loaded for the language `#1'. Using the pattern for}%
\typeout{** the default language instead.}%
\else
\language=\csname l@#1\endcsname
\fi
#2}}
\providecommand{\BIBdecl}{\relax}
\BIBdecl

\bibitem{1}
G.~Qu, Q.~Chen, W.~Wei \emph{et~al.}, ``Mobile edge intelligence for large
  language models: {A} contemporary survey,'' \emph{IEEE Commun. Surveys
  Tuts.}, vol.~27, no.~6, pp. 3820--3860, 2025.

\bibitem{2}
D.~Wen, Y.~Zhou, X.~Li \emph{et~al.}, ``A survey on integrated sensing,
  communication, and computation,'' \emph{IEEE Commun. Surveys Tuts.}, vol.~27,
  no.~5, pp. 3058--3098, 2025.

\bibitem{3}
\BIBentryALTinterwordspacing
ITU-R, ``Framework and overall objectives of the future development of {IMT}
  for 2030 and beyond,'' Nov. 2023. [Online]. Available:
  \url{https://www.itu.int/en/ITU-R/study-groups/rsg5/rwp5d/imt-2030/Pages/default.aspx}
\BIBentrySTDinterwordspacing

\bibitem{4}
J.~Yao, W.~Xu, G.~Zhu \emph{et~al.}, ``Energy-efficient edge inference in
  integrated sensing, communication, and computation networks,'' \emph{IEEE J.
  Select. Areas Commun.}, vol.~43, no.~10, pp. 3580--3595, 2025.

\bibitem{5}
Y.~Mao, X.~Yu, K.~Huang \emph{et~al.}, ``Green edge {AI}: {A} contemporary
  survey,'' \emph{Proc. IEEE}, vol. 112, no.~7, pp. 880--911, 2024.

\bibitem{6}
Z.~Wang, Y.~Zhao, Y.~Zhou \emph{et~al.}, ``Over-the-air computation for {6G}:
  {F}oundations, technologies, and applications,'' \emph{IEEE Internet Things
  J.}, vol.~11, no.~14, pp. 24\,634--24\,658, 2024.

\bibitem{7}
Y.~Fu, P.~Qin, G.~Tang \emph{et~al.}, ``Joint design of sensing, communication,
  and computation for multi-{UAV}-enabled over-the-air federated learning,''
  \emph{IEEE Trans. Veh. Technol.}, vol.~74, no.~9, pp. 13\,909--13\,924, 2025.

\bibitem{Du2024Distributed}
J.~Du, T.~Lin, C.~Jiang \emph{et~al.}, ``Distributed foundation models for
  multi-modal learning in {6G} wireless networks,'' \emph{IEEE Wireless
  Commun.}, vol.~31, no.~3, pp. 20--30, 2024.

\bibitem{8}
Z.~Liu, Q.~Lan, A.~E. Kalør \emph{et~al.}, ``Over-the-air multi-view pooling
  for distributed sensing,'' \emph{IEEE Trans. Wireless Commun.}, vol.~23,
  no.~7, pp. 7652--7667, 2024.

\bibitem{9}
X.~Chen, K.~B. Letaief, and K.~Huang, ``On the view-and-channel aggregation
  gain in integrated sensing and edge {AI},'' \emph{IEEE J. Select. Areas
  Commun.}, vol.~42, no.~9, pp. 2292--2305, 2024.

\bibitem{10}
X.~Shi, J.~Du, J.~Wang \emph{et~al.}, ``Beamforming design for massive
  {MIMO}-aided over-the-air computation: {A} mutual information perspective,''
  \emph{IEEE Trans. Wireless Commun.}, vol.~23, no.~10, pp. 14\,335--14\,349,
  2024.

\bibitem{11}
Z.~Liu, Q.~Lan, and K.~Huang, ``Over-the-air fusion of sparse spatial features
  for integrated sensing and edge {AI} over broadband channels,'' \emph{IEEE
  Trans. Wireless Commun.}, vol.~24, no.~4, pp. 2999--3013, 2025.

\bibitem{12}
Y.~Fu, P.~Qin, Y.~Wang \emph{et~al.}, ``Over-the-air edge inference for
  low-altitude airspace: {G}enerative {AI}-aided multi-task batching and
  beamforming design,'' \emph{IEEE Trans. Commun.}, vol.~73, no.~10, pp.
  9013--9027, 2025.

\bibitem{13}
G.~Zhu, Y.~Du, D.~Gündüz \emph{et~al.}, ``One-bit over-the-air aggregation
  for communication-efficient federated edge learning: {D}esign and convergence
  analysis,'' \emph{IEEE Trans. Wireless Commun.}, vol.~20, no.~3, pp.
  2120--2135, 2021.

\bibitem{14}
L.~You, X.~Zhao, R.~Cao \emph{et~al.}, ``Broadband digital over-the-air
  computation for wireless federated edge learning,'' \emph{IEEE Trans. Mobile
  Comput.}, vol.~23, no.~5, pp. 5212--5228, 2024.

\bibitem{15}
A.~Şahin, ``Over-the-air computation based on balanced number systems for
  federated edge learning,'' \emph{IEEE Trans. Wireless Commun.}, vol.~23,
  no.~5, pp. 4564--4579, 2024.

\bibitem{16}
J.~Liu, Y.~Gong, and K.~Huang, ``Digital over-the-air computation: {A}chieving
  high reliability via bit-slicing,'' \emph{IEEE Trans. Wireless Commun.},
  vol.~24, no.~5, pp. 4101--4114, 2025.

\bibitem{17}
L.~Qiao, Z.~Gao, M.~Boloursaz~Mashhadi \emph{et~al.}, ``Massive digital
  over-the-air computation for communication-efficient federated edge
  learning,'' \emph{IEEE J. Select. Areas Commun.}, vol.~42, no.~11, pp.
  3078--3094, 2024.

\bibitem{18}
Y.~Fu, P.~Qin, J.~Zhang \emph{et~al.}, ``Joint {AI} inference and target
  tracking at network edge: {A} hybrid offline-online design for {UAV}-enabled
  network,'' \emph{IEEE Trans. Wireless Commun.}, vol.~23, no.~12, pp.
  17\,959--17\,973, 2024.

\bibitem{19}
Z.~Lin, L.~Yang, Y.~Gong \emph{et~al.}, ``Semantic-topology preserving
  quantization of word embeddings for human-to-machine communications,''
  \emph{IEEE Trans. Commun.}, vol.~73, no.~4, pp. 2401--2415, 2025.

\bibitem{20}
X.~Zhai, G.~Han, and Y.~Cai, ``Robust design of {RIS}-assisted over-the-air
  computation in the face of long-term and short-term imperfections,''
  \emph{IEEE Trans. Veh. Technol.}, vol.~73, no.~11, pp. 16\,817--16\,831,
  2024.

\bibitem{21}
X.~Zhai, G.~Han, Y.~Cai \emph{et~al.}, ``Joint beamforming aided over-the-air
  computation systems relying on both {BS}-side and user-side reconfigurable
  intelligent surfaces,'' \emph{IEEE Trans. Wireless Commun.}, vol.~21, no.~12,
  pp. 10\,766--10\,779, 2022.

\bibitem{22}
B.~Wei, P.~Zhang, and Q.~Zhang, ``Active reconfigurable intelligent
  surface-aided over-the-air computation networks,'' \emph{IEEE Wireless
  Commun. Lett.}, vol.~13, no.~4, pp. 1148--1152, 2024.

\bibitem{23}
X.~Zhai, G.~Han, Y.~Cai \emph{et~al.}, ``Simultaneously transmitting and
  reflecting ({STAR}) {RIS} assisted over-the-air computation systems,''
  \emph{IEEE Trans. Commun.}, vol.~71, no.~3, pp. 1309--1322, 2023.

\bibitem{24}
A.~Zheng, W.~Ni, W.~Wang \emph{et~al.}, ``Multi-functional {RIS} for
  distributed over-the-air computation in base station free environments,''
  \emph{IEEE Trans. Commun.}, vol.~73, no.~10, pp. 8840--8855, 2025.

\bibitem{25}
Y.~Ju, S.~Gong, H.~Liu \emph{et~al.}, ``Beamforming optimization for hybrid
  active-passive {RIS} assisted wireless communications: {A} rate-maximization
  perspective,'' \emph{IEEE Trans. Commun.}, vol.~72, no.~9, pp. 5428--5442,
  2024.

\bibitem{26}
Z.~Kang, C.~You, and R.~Zhang, ``Active-passive {IRS} aided wireless
  communication: {N}ew hybrid architecture and elements allocation
  optimization,'' \emph{IEEE Trans. Wireless Commun.}, vol.~23, no.~4, pp.
  3450--3464, 2024.

\bibitem{Chen2024Physical}
Z.~Chen, Y.~Guo, P.~Zhang \emph{et~al.}, ``Physical layer security improvement
  for hybrid {RIS}-assisted {MIMO} communications,'' \emph{IEEE Commun. Lett.},
  vol.~28, no.~11, pp. 2493--2497, 2024.

\bibitem{hardware1}
Q.~Li, M.~El-Hajjar, C.~Xu \emph{et~al.}, ``Stacked intelligent metasurfaces
  for holographic {MIMO}-aided cell-free networks,'' \emph{IEEE Trans.
  Commun.}, vol.~72, no.~11, pp. 7139--7151, 2024.

\bibitem{hardware2}
Q.~Li, M.~El-Hajjar, K.~Cao \emph{et~al.}, ``Holographic metasurface-based
  beamforming for multi-altitude {LEO} satellite networks,'' \emph{IEEE Trans.
  Wireless Commun.}, vol.~24, no.~4, pp. 3103--3116, 2025.

\bibitem{hardware3}
Q.~Li, M.~El-Hajjar, Y.~Sun \emph{et~al.}, ``Performance analysis of
  reconfigurable holographic surfaces in the near-field scenario of cell-free
  networks under hardware impairments,'' \emph{IEEE Trans. Wireless Commun.},
  vol.~23, no.~9, pp. 11\,972--11\,984, 2024.

\bibitem{33}
Y.~Lin, F.~Wang, X.~Zhang \emph{et~al.}, ``Joint mode selection and beamforming
  designs for hybrid-{RIS}-assisted {ISAC} systems,'' \emph{IEEE Wireless
  Commun. Lett.}, vol.~14, no.~6, pp. 1718--1722, 2025.

\bibitem{28}
Y.~Du, S.~Yang, and K.~Huang, ``High-dimensional stochastic gradient
  quantization for communication-efficient edge learning,'' \emph{IEEE Trans.
  Signal Processing}, vol.~68, pp. 2128--2142, 2020.

\bibitem{27}
W.~Dai, Y.~Liu, and B.~Rider, ``Quantization bounds on {G}rassmann manifolds
  and applications to {MIMO} communications,'' \emph{IEEE Trans. Inform.
  Theory}, vol.~54, no.~3, pp. 1108--1123, 2008.

\bibitem{29}
\BIBentryALTinterwordspacing
{3GPP, document TS 38.300 V15.3.1}, ``{5G} {NR} overall description
  {S}tage-2,'' Oct. 2018. [Online]. Available:
  \url{https://www.3gpp.org/ftp/Specs/archive/38_series/38.300/}
\BIBentrySTDinterwordspacing

\bibitem{30}
Z.~Zhuang, D.~Wen, Y.~Shi \emph{et~al.}, ``Integrated
  sensing-communication-computation for over-the-air edge {AI} inference,''
  \emph{IEEE Trans. Wireless Commun.}, vol.~23, no.~4, pp. 3205--3220, 2024.

\bibitem{32}
T.~Blumensath and M.~E. Davies, ``Stagewise weak gradient pursuits,''
  \emph{IEEE Trans. Signal Processing}, vol.~57, no.~11, pp. 4333--4346, 2009.

\bibitem{34}
H.~Shen, W.~Xu, S.~Gong \emph{et~al.}, ``Beamforming optimization for
  {IRS}-aided communications with transceiver hardware impairments,''
  \emph{IEEE Trans. Commun.}, vol.~69, no.~2, pp. 1214--1227, 2021.

\bibitem{35}
\BIBentryALTinterwordspacing
J.-C. Su, M.~Gadelha, R.~Wang \emph{et~al.}, ``A deeper look at {3D} shape
  classifiers,'' 2018. [Online]. Available:
  \url{https://arxiv.org/abs/1809.02560}
\BIBentrySTDinterwordspacing

\bibitem{36}
L.~Qiao, J.~Zhang, Z.~Gao \emph{et~al.}, ``Joint activity and blind information
  detection for {UAV}-assisted massive {I}o{T} access,'' \emph{IEEE J. Select.
  Areas Commun.}, vol.~40, no.~5, pp. 1489--1508, 2022.

\end{thebibliography}
   
\end{document}